\documentclass[11pt,a4paper,final]{iopart}
\pdfoutput=1
\usepackage{graphicx}
\usepackage[breaklinks=true,colorlinks=true,linkcolor=blue,urlcolor=blue,citecolor=blue]{hyperref}
\usepackage{bbm}
\usepackage{amssymb}
\usepackage{dcolumn} % needed for some tables
\PassOptionsToPackage{hyphens}{url}
\usepackage{hyperref}

\newcommand{\rev}[1]{{\color{black}#1}}
\newcommand{\ep}{\varepsilon}
\newcommand{\lm}{\lambda}

\newcommand{\be}{\begin{equation}}
\newcommand{\ee}{\end{equation}}
\def\ba{\begin{aligned}}
\def\ea{\end{aligned}}
\newcommand{\bea}{\begin{eqnarray}}
\newcommand{\eea}{\end{eqnarray}}

\newcommand{\la}{\left\langle}
\newcommand{\ra}{\right\rangle}

\renewcommand{\vec}[1]{{\bf #1}}

\begin{document}
\title[Crooks and Gallavotti-Cohen theorems in driven classical Markovian systems]
%      {Relations between Crooks and Gallavotti-Cohen fluctuation relations in driven classical Markovian systems}
      {Time-reversal symmetric Crooks and Gallavotti-Cohen fluctuation relations in driven classical Markovian systems}

\author{A.~Mandaiya~$^{1}$, I.~M.~Khaymovich~$^{2,3}$}
 \address{$^1$ Department of Physics,  Cornell University, Ithaca, NY 14853, USA}
 \address{$^2$ Max-Planck-Institut f\"ur Physik komplexer Systeme, N\"othnitzer Stra{\ss}e~38, 01187-Dresden, Germany }
 \address{$^3$ Institute for Physics of Microstructures, Russian Academy of Sciences, 603950 Nizhny Novgorod, GSP-105, Russia}
\ead{ivan.khaymovich@pks.mpg.de}
\vspace{10pt}
\begin{indented}
\item[]December 2018
\end{indented}

\begin{abstract}
In this paper, we address an important question of the relationship between fluctuation theorems for
the dissipated work $W_{d} = W-\Delta F$
with general finite-time (like Jarzynski equality and Crooks relation) and infinite-time (like Gallavotti-Cohen theorem) drive protocols and their time-reversal symmetric versions.
The relations between these kinds of fluctuation relations are uncovered based on the examples of a classical Markovian $N$-level system.
Further consequences of these relations are discussed with respect to the possible experimental verifications.
\end{abstract}

%
% Uncomment for keywords
\vspace{2pc}

% Uncomment for Submitted to journal title message
\submitto{\JSTAT}

% Uncomment if a separate title page is required
\maketitle
%
% For two-column output uncomment the next line and choose [10pt] rather than [12pt] in the \documentclass declaration
%\ioptwocol
%

\maketitle

%%%%%%%%%%%%%%%%%%%%%%%%%%%%%%%%%%%%%%%%%%%%%%%%%%%%%%%%%%%%%%%%%%%%%%%%%%%%%%%%%%%%
\section{Introduction}
%%%%%%%%%%%%%%%%%%%%%%%%%%%%%%%%%%%%%%%%%%%%%%%%%%%%%%%%%%%%%%%%%%%%%%%%%%%%%%%%%%%%

For the last decades, since $1970$s~\cite{BK} when the first fluctuation theorems generalizing the second law of thermodynamics were formulated~(see review~\cite{BK-review} and references therein),
there have been discovered many variants of fluctuation relations spreading from the ones for the heat and environmental entropy production in the static conditions, either in non-equilibrium steady state (NESS)~\cite{Evans-Cohen1993,GC1995,Kurchan1998,Lebowitz-Spohn1999} or during relaxation to equilibrium~\cite{Evans-Searles1994}, to the well-known Jarzynski equality~\cite{Jarzynski1997} and Crooks relation~\cite{Crooks1999} written for the work dissipated in the system under a finite-time drive.
Some work has been done on their generalizations for the periodic drive~\cite{Tietz-Seifert2006,Shargel-Chou2009}
and for the stochastic entropy production~\cite{Seifert2005,Garcia-Garcia2010} which are less known (please see~\cite{Seifert2012} for the extensive review).

Experimental verifications of different kinds of fluctuation relations has been initiated by measurements in biological systems~\cite{Liphardt2002} and then done in various different classical systems, such as mechanical~\cite{Evans2002,Ciliberto2005,Hoang2018}, biological~\cite{Bustamante2004,Collin2005}, and condensed matter systems both in contact with equilibrium~\cite{Tietz-Seifert2006,Ciliberto2005_Resistor,Schuler-Seifert2005,Saira2012,Roldan-Ciliberto2015,Hofmann2015} and non-equilibrium~\cite{Koski2013} environment.
In most of these systems, thermodynamic variables (work, heat, or entropy) has been extracted indirectly via the measurement of the microscopic state of the system (a position of the bead in a laser tweezer, an instantaneous angular deflection of the rotation pendulum, a charge state of a Coulomb-blockaded device and so on).
Direct measurements of the heat or work especially in quantum systems~\cite{Serra2014,Nakamura2010} have not been done yet, but many efforts have been undertaken, especially in the most stable Coulomb-blockaded devices~\cite{Gasparinetti2015,Feshchenko2015,Koski2015_AutoMD,Saira2016,Zgirski2018,Wang2018}.

Recently fluctuation relations have been also generalized to the case of a feedback-controlled systems~\cite{Sagawa-Ueda2010,Sagawa-Ueda2012} including recent ones like~\cite{Potts2018} which has opened a path to understand the paradox of Maxwell's Demon from the Landauer's principle~\cite{Landauer1961, Landauer1988} and
verify these predictions  experimentally in finite-time protocols~\cite{Toyabe2010,Roldan2014,Koski2014_MD_PNAS,Koski2014_MD_PRL,Vidrighin2016,Ribezzi,Chida2015,Wagner2016} and even in the steady-state conditions in the autonomous realization of a Szilard engine~\cite{Koski2015_AutoMD} with the direct measurements of the effect of the feedback both on the system and demon's temperatures, giving a direct access to the demon's thermodynamics (see recent reviews for the details~\cite{Pekola2015_NatPhys_Review,Pekola2019_AnnRev}).

Recent theoretical progress has already provided a more detailed information about properties of the large fluctuations both in the stochastic entropy production~\cite{Chetrite_Gupta,Neri_PRX,Pigolotti_PRL,Neri2019} in NESS (with experimental verification in~\cite{Singh_Sinf}, including quantum systems~\cite{Manzano2019}) and in the heat in driven systems~\cite{Chetrite_Q-Martingale} basing on a Martingale theory.

Despite the impressive progress in the understanding of physics of fluctuations until now, the relations between fluctuation theorems in the systems under finite-time drives and in NESS (or periodic NESS) has been only barely studied.
For example, in the work~\cite{VerleyLacoste2012} the importance of initial conditions for finite-time
fluctuation theorems in NESS comparing to their asymptotic long-time counterparts has been discussed.
In this paper, we address the important and demanding question of these relations between fluctuation theorems for driven systems on an example of a classical Markovian $N$-level system.

The paper is organized as follows.
In Sec.~\ref{Sec:Model}, we describe the model, overview briefly main fluctuation theorems, and formulate the main question in the focus.
Section~\ref{Sec:P(W)} gives the standard method of the calculation~\cite{Sekimoto} of the probability distribution of the dissipated work in a driven system and provides main equations used further.
In Sec.~\ref{Sec:TRS_and_beyond}, we derive the conditions when the fluctuation relations can be written in the time-reversal symmetric case and extend the class of drive protocols for which these conditions are satisfied.
Section~\ref{Sec:TLS} is devoted to the consideration of the relations of finite-time and periodic-NESS fluctuation relations in a two-level system, where we provide an exact correspondence between finite-time and periodic-NESS fluctuation theorems.
Section~\ref{Sec:Conclusion} concludes our paper.

%%%%%%%%%%%%%%%%%%%%%%%%%%%%%%%%%%%%%%%%%%%%%%%%%%%%%%%%%%%%%%%%%%%%%%%%%%%%%%%%%%%%
\section{Model and definitions}\label{Sec:Model}
%%%%%%%%%%%%%%%%%%%%%%%%%%%%%%%%%%%%%%%%%%%%%%%%%%%%%%%%%%%%%%%%%%%%%%%%%%%%%%%%%%%%
In this section, we consider a Markovian $N$-level system.
The formalism of this section is standard and for more details please address, e.g., the book~\cite{Sekimoto}.
The system in focus is
characterized by the energy levels $E_n(\lm)$, $n=\overline{0,N-1}$,
and subjected to the drive via a time-dependent control parameter $\lambda(t)$.
The system is placed in contact with a bath with a certain inverse temperature, $\beta$.
The Markovian dynamics of the considered system is described by the standard rate equations  written in the matrix form
\be\label{eq:rate_eqs}
\frac{d}{dt}\left|\vec{p}(t)\ra = \hat\Gamma(\lm(t)) \left|\vec{p}(t)\ra \
\ee
for the vector $\left|\vec{p}(t)\ra = (p_0,\ldots, p_{N-1})$ of probabilities $p_n(t)$ of the system to be in the state $n$ at a certain time
instant $t$.
In the main part of the paper, for simplicity, we consider the case when time-dependent incoming rates $\Gamma_{n,n'}(\lm(t))$ from states $n'$ to a certain state $n$
satisfy the local detailed balance (LDB) condition
\be\label{eq:LDB}
\Gamma_{n',n}(\lm(t)) = \Gamma_{n,n'}(\lm(t))e^{\beta[E_n(\lm(t))-E_{n'}(\lm(t))]} \ .
\ee
The normalization condition for the probability distribution $\la \vec{1}|\vec{p}\ra\equiv\sum_{n=0}^{N-1} p_n(t) = 1$, with $\left|\vec{1}\ra = (1,\ldots,1)$, is conserved by rate
equations as the overall escape rate from the state $n$ is
$\Gamma_{n,n} = \sum_{n'\ne n} \Gamma_{n',n}$.
Here and further, we put the Boltzmann's constant to be unity, i.e., $k_{B}=1$ and measure temperature in energy units.
The initial distribution $p_n(0)$ of the system is considered to be equilibrium
\be\label{eq:p_Eq}
\left|\vec{p}_{eq}(\lm(0))\ra = e^{\beta [F(\lm(0))-\hat E(\lm(0))]}\left|\vec{1}\ra \ ,
\ee
where $E_{n,n'}(\lm)=\delta_{n,n'}E_n(\lm)$ is a diagonal matrix of system's energy levels and $\beta F (\lm) = - \ln\sum_n e^{-\beta E_n(\lm)}$ is
the free energy of the system at a certain value of $\lm(t)=\lm$.

\begin{figure}[t!]
\hfill\includegraphics[width=\textwidth]{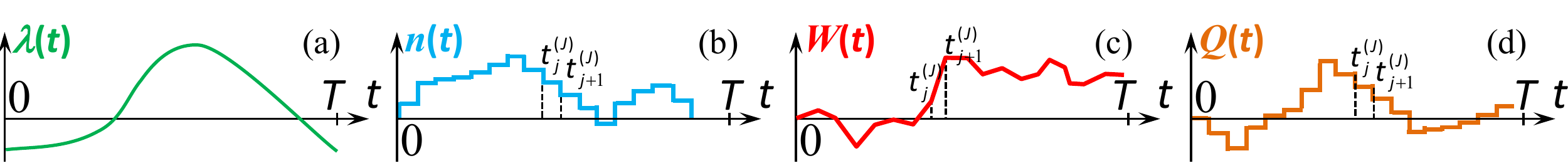}
\caption{Sketch of (a)~the general cyclic drive protocol $\lambda(t)$ (green line),
(b)~a single piecewise constant trajectory $n(t)$ of the system state (blue line), which jumps at time instants $t_j^{(J)}$ from state $n_{j-1}$ to $n_j$, and corresponding (c)~work (red line) and (d)~heat (orange line) on this trajectory.
} \label{Fig0_lambda_n_W_Q}
\end{figure}

%\subsection{Stochastic work, heat, and entropy production}
The first law of thermodynamics $dE_{n(t)}(\lm(t)) = \delta W + \delta Q$,
written in terms of the system internal energy $E_{n(t)}(\lm(t))$
gives the definitions of the work performed to the system
\be\label{eq:work_def}
W = \int_0^T \left.\frac{\partial E_n}{\partial \lm}\right|_{n(t)} \frac{d\lm}{dt} dt = \sum_j \left[E_{n_j}(t_{j+1}^{(J)})-E_{n_{j}}(t_j^{(J)})\right] \ ,
\ee
and the heat dissipated to the bath
\be
\label{eq:heat_def}
Q = -\int_0^T \left.\frac{\partial E_n}{\partial n}\right|_{\lm(t)}\frac{d n}{dt} dt = \sum_j \left[E_{n_j}(t_j^{(J)})-E_{n_{j-1}}(t_j^{(J)})\right] \ .
\ee
being the changes of $E_{n(t)}(\lm(t))$ with respect to the control parameter $\lm(t)$ and the system state $n(t)$, respectively, see Fig.~\ref{Fig0_lambda_n_W_Q}.
Here and further, we consider the evolution of the system's state $n(t)$ as a set of jumps from $n_{j-1}$ to $n_j$ occurred at time instant $t_j^{(J)}$, see Fig.~\ref{Fig0_lambda_n_W_Q}(b).

%%%%%%%%%%%%%%%%%%%%%%%%%%%%%%%%%%%%%%%%%%%%%
%%%%%%%%%%%%%%%%%%%%%%%%%%%%%%%%%%%%%

For driven systems which obey LDB~(\ref{eq:LDB}) under a finite-time drive $\lm(t)$, $0\leq t\leq T$,
and start from the equilibrium distribution~(\ref{eq:p_Eq}),
the probability distribution of work is characterized by
the Jarzynski equality~\cite{Jarzynski1997}
\be\label{eq:JE_W}
\la e^{-\beta W}\ra = e^{-\beta \Delta F}
\ee
and the Crooks relation~\cite{Crooks1999}
\be\label{eq:Crooks_W}
P(W)/\bar P(-W) = e^{\beta (W-\Delta F)} \ .
\ee
Here, the averaging $\la \ldots\ra$ is performed over all microscopic realizations of the system and the bath during the protocol $\lm (t)$,
$\bar P(W)$ denotes the probability distribution of work in the time-reversed drive protocol $\lambda(T-t)$.
%For the time-reversal symmetric (TRS) drive, $\lm(T-t)=\lm(t)$, the latter coincides with the distribution in the forward-time protocol, $\bar P(W) = P(W)$,
%posing a certain symmetry on the distribution $P(W)$ and
%quite unexpectedly providing the ground for several intriguing results such as
%the straightforward calculations of the first-passage-time distribution from the fixed-time distribution~\cite{Singh_Thesis,Singh_1st_passage} and
%the surprising analogy of the work statistics with the multifractality of the wavefunctions close to the Anderson localization
%transition~\cite{MF_analogy}.

To lift the equilibrium condition on the initial distribution~(\ref{eq:p_Eq}), one has to consider the large-deviation version~\cite{Touchette2009}
(sometimes called weak version~\cite{VerleyLacoste2012}) of Crook's
relation~\cite{GC1995}
for the asymptotic long-time limit
\be\label{eq:GC_W}
\lim_{t\to\infty} \frac{1}{t} \ln \frac{P(W=wt)}{\bar P(-W=-wt)} = \beta w \ ,
\ee
where $w = W/t$ is an intensive parameter of work. The free energy rate $\Delta F/t$ is negligible in infinite-time limit as the free energy difference $\Delta F$ is bounded.
Note that the analogous large-deviation Crook's relation can be written for the heat rate $q=Q/t$ as the internal energy change $\Delta
E_n(\lambda)$ is bounded for all
finite values of $\lm$.
Further, for simplicity, we will omit the explicit dependence of $E_n$ and $p_{eq,n}$ %and $F$
on $\lm(t)$, keeping only $t$ as an argument.

We complete the introductory part of the paper by considering briefly the stochastic entropy productions.
Stochastic entropy production of the environment, $\Delta s_m$, generalizes the concept of the heat for the systems violating the LDB
condition~(\ref{eq:LDB}).
Indeed, like heat $Q$~(\ref{eq:heat_def}),
this quantity sums the jumps $\Delta s_m = \sum_i \Delta s_{m,n_{j-1}\to n_j}(t_j^{(J)})$
occurring as soon as the state $n(t)$ of the system changes
(from $n_{j-1}$ to $n_j$ occurred at time instant $t_j^{(J)}$), however, the size of each jump
\be\label{eq:s_m_def}
\Delta s_{m,n\to n'}(t) = \ln \frac{\Gamma_{n',n}(t)}{\Gamma_{n,n'}(t)} \
\ee
coincides with the one $\Delta Q_{n\to n'}(t)\equiv \left[E_{n'}(t)-E_{n}(t)\right]$ of $Q$ multiplied by $\beta$ only when the system obeys
LDB~(\ref{eq:LDB}).

The analogue of the dissipated work, $W-\Delta F$, for this case is the total entropy production introduced in~\cite{Seifert2005}. It is given by the sum
\be\label{eq:s_tot_def}
\Delta s_{tot} = \Delta s_m + \Delta s_{sys}
\ee
of the environmental $\Delta s_m$ and system entropy change $\Delta s_{sys} = s_{sys}(T) - s_{sys}(0)$,
where
\be\label{eq:s_sys_def}
s_{sys} = -\ln p_{n(t)}(t) \
\ee
is the stochastic analogue of the Shannon's entropy given by $\la s_{sys}\ra = -\sum_n p_n \ln p_n$.
The main property of the stochastic total entropy production $\Delta s_{tot}$ is that it satisfies the generalized Jarzynski equality and the Crooks
relations~\cite{Seifert2005}, called sometimes the integral and \rev{detailed fluctuation relations (DFR)}, respectively~\cite{Seifert2012},
\bea\label{eq:JE_S_tot}
\langle e^{-\Delta s_{tot}}\rangle &=& 1 \ ,\\
\label{eq:Crooks_S_tot}
P(\Delta s_{tot})/\bar P(-\Delta s_{tot}) &=& e^{\Delta s_{tot}} \ .
\eea
These fluctuation relations work beyond LDB condition and for any initial distribution.
However, the price paid for lifting of LDB and the equilibrium initial distribution is that $\Delta s_{tot}$
depends not only on a single trajectory realized by a system, but also on its instantaneous probability distribution via $s_{sys}(t)$.
However recently it have been found that certain decompositions of the stochastic entropy production
provide the representation of $\Delta s_{tot}$ on a single trajectory in terms of physical observables like work and particle current for some given initial conditions~\cite{CuetaraEsposito2014,RaoEsposito2018}.

The large-deviation variant of \rev{DFR}~(\ref{eq:Crooks_S_tot})
\be\label{eq:GC_S_tot}
\lim_{t\to\infty} \frac{1}{t} \ln \frac{P(\Delta s_{tot}=\sigma t)}{\bar P(-\Delta s_{tot}=-\sigma t)} = \sigma \ ,
\ee
has been originally written in the paper~\cite{GC1995} for the environmental entropy
in the system in the non-equilibrium steady state (NESS), as the system entropy production is intensive quantity (as well as the internal and free energies).
Note that the large-deviation Crooks relation for the work in NESS conditions is trivial as the control parameter $\lm$ is constant and the work is
zero.
To avoid this triviality, further, we consider the periodic-drive condition inferring periodic-NESS~\cite{Talkner1999}.
Thus, the free energy difference can be omitted in both Eqs.~(\ref{eq:Crooks_W}, \ref{eq:GC_W}).

Obviously in all considered variants~(\ref{eq:Crooks_W}, \ref{eq:GC_W}, \ref{eq:Crooks_S_tot}, \ref{eq:GC_S_tot}) of \rev{DFR} the probability distribution
$\bar P$ in the denominator coincides with the one in the numerator $P$
provided the drive protocol is time-reversal symmetric (TRS), $\lm(T-t)=\lm(t)$~\footnote{However, in the large-deviation versions it is enough that the drive would be symmetric with respect to an arbitrary finite time shift, see, e.g., Fig.~\ref{Fig:2step}.}
\bea\label{eq:Crooks_W_TRS}
P(W)/P(-W) &=& e^{\beta (W-\Delta F)} \ , \\
\label{eq:Crooks_S_tot_TRS}
P(\Delta s_{tot})/P(-\Delta s_{tot}) &=& e^{\Delta s_{tot}} \ , \\
\label{eq:GC_W_TRS}
\lim_{t\to\infty} t^{-1} \ln \left[{P(wt)}/{P(-wt)}\right] &=& \beta w \ , \\
\label{eq:GC_S_tot_TRS}
\lim_{t\to\infty} t^{-1} \ln \left[{P(\sigma t)}/{P(-\sigma t)}\right] &=& \sigma \ .
\eea
This poses a certain symmetry restrictions on the distribution $P$ and opens an intriguing possibility for the direct calculations of first-passage-time distribution for considered variables from their
distributions at fixed time~\cite{Singh_Thesis,Singh_1st_passage}.
Another issue emerging from the relations (\ref{eq:Crooks_W_TRS}~--~\ref{eq:GC_S_tot_TRS}) is the surprising analogy of the work statistics with the multifractality of
the wavefunctions close to the Anderson localization transition considered in~\cite{MF_analogy}.

Both for the dissipated work and for the total entropy production an important question arises:
What is the relation between large deviation and finite-time versions of Crooks relations?
In particular, %one can ask whether the TRS is necessary to formulate Crook-like relations for the only distribution function $P$ and if not,
what are the requirements on the drive beyond TRS for a system to obey
Crook-like relations for the only distribution function $P$ and what are the
relations between these requirements for finite-time protocol and periodic-NESS?

To address all these questions in the next section, we describe the standard method to calculate the
probability distributions by writing the rate equations for the generating functions
and focus mostly on the dissipated work normalized to the temperature $w_d = \beta(W-\Delta F)$ as a variable of interest.
Please see~\ref{App_Sec:gen_func} for the general method given, e.g., in the book~\cite{Sekimoto} for other thermodynamics variables mentioned above.

%%%%%%%%%%%%%%%%%%%%%%%%%%%%%%%%%%%%%%%%%%%%%%%%%%%%%%%%%%%%%%%%%%%%%%%%%%%%%%%%%%%%
\section{Calculation of $P(W-\Delta F)$}\label{Sec:P(W)}
%%%%%%%%%%%%%%%%%%%%%%%%%%%%%%%%%%%%%%%%%%%%%%%%%%%%%%%%%%%%%%%%%%%%%%%%%%%%%%%%%%%%
In order to write the rate equation of the form similar to~(\ref{eq:rate_eqs}) one should consider
the $n$-resolved distribution function $\left|\vec{P}(w_d)\ra = (P_0(w_d),\ldots,P_{N-1}(w_d))$, with the components defined as
\be
P_n(\beta(W-\Delta F)=w_d) = \langle \delta(\beta W-\beta \Delta F-w_d)\delta_{n,n(t)}\rangle \ , %= \frac{1}{2\pi}\int G^W_{q,n} e^{-iq W} dq
\ee
because the probability distribution itself $P(w_d) = \la \vec{1}|\vec{P}(w_d)\ra\equiv \sum_n P_n(w_d)$ does not determine explicitly
the system state $n(t)$.
To simplify the derivation even further we go to the Laplace transform of $\left|\vec{P}(w_d)\ra$ being
%In general it is difficult to write the rate equation for the $n$-resolved distribution function $\left|\vec{P}(w_d)\ra$ itself, but one can do it for
the $n$-resolved generating function~\footnote{Rate equations for the $n$-resolved distribution function $\left|\vec{P}(w_d)\ra$ itself are given in~\ref{App_Sec:gen_func} or~\cite{Esposito2013}.}
 %defined as the Laplace transform of the latter
\be\label{eq:G_q_vec}
\left|\vec{G}_{q}\ra = \int \left|\vec{P}(w_d)\ra e^{-q w_d} dw_d \ .
\ee

Using the standard trajectory representation of the jump Markov processes widely used in the full counting statistics (see,
e.g.,~\cite{Bagrets_Nazarov2003}),
one can derive the rate equations of the form of (\ref{eq:rate_eqs})
\be\label{eq:rate_eqs_G_q,n}
\frac{d}{dt}\left|\vec{G}_{q}(t)\ra = \hat\Gamma^{(q)}(t)\left|\vec{G}_{q}(t)\ra \ ,
\ee
with the modified rate matrix $\hat\Gamma^{(q)}(t)$, and the initial condition $\left|\vec{G}_{q}(0)\ra=\left|\vec{p}(0)\ra$ provided
$w_d(0)=0$.
For the dissipated work which rate $\dot w_{d,n}(t)$ is a deterministic function of $n(t)$ only the escape rates should be
modified
\be\label{eq:Gamma_n->n'_W}
\Gamma^{(q)}_{n,n}(t) = \Gamma_{n,n}(t) + q  \dot w_{d,n}(t) \ ,
\ee
with $\dot w_{d,n} = {\partial e_{n}}/{\partial t}|_{n(t)}$ and $e_{n}(t) = \beta(E_n(t)-F(t))$.
%Please see~\ref{App_Sec:gen_func} \rev{or the book~\cite{Sekimoto}} for the general case of modified rates~(\ref{eq:Gamma_n->n'_W}) in the equations~(\ref{eq:rate_eqs_G_q,n}) for other thermodynamics variables mentioned above.
Note that unlike Eq.~(\ref{eq:rate_eqs}) the latter equation does not conserve normalization condition as $\Gamma^{(q)}_{n,n} \ne
\sum_{n'\ne n} \Gamma^{(q)}_{n',n}$.

The probability distribution of $w_d$
\be\label{eq:P(x)}
P(w_d) = \frac{1}{2\pi i}\lim_{Q\to\infty}\int_{\chi-iQ}^{\chi+iQ} G_{q}(t) e^{q w_d} d q \
\ee
is given by the inverse Laplace transform of the generating function
\be\label{eq:G_q}
G_{q}(t) =\la \vec{1}|\vec{G}_q(t)\ra\equiv \sum_n G_{q,n}(t) \ .
\ee
The parameter $\chi$ is greater than real part of all singularities of
$G_{q}(t)$ as a function of $q$.

\begin{figure}[t!]
\hfill\includegraphics[width=\textwidth]{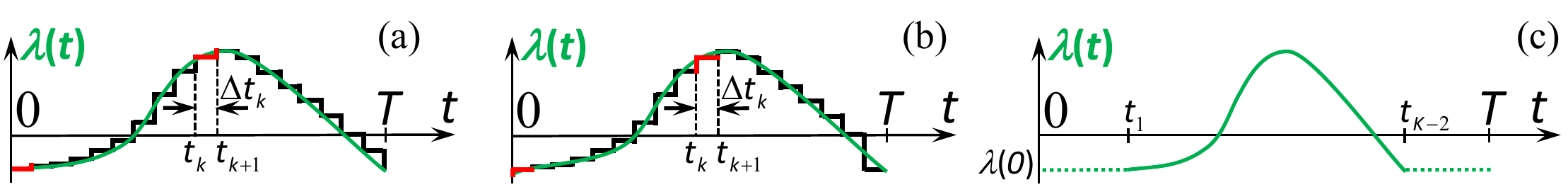}
\caption{(a, b) Sketch of the general cyclic drive protocol $\lambda(t)$ (green lines) and its time discretized forms (black lines)
(a) with plateaux followed by jumps and (b) jumps followed by plateaux (both emphasized in red);
(c) Modified cyclic drive protocol with the zeroth, $0=t_0<t<t_1$, and last, $t_{K-2}<t<t_{K-1}=T$, intervals of a constant drive (dashed lines).
The latter intervals do not contribute to the work generating function $G_q$ as the constant drive does not change work in equilibrium ($0=t_0<t<t_1$) or relaxation ($t_{K-2}<t<t_{K-1}=T$) part.
} \label{Fig1_step_drive}
\end{figure}

The generating function (\ref{eq:G_q}),
both for finite-time and periodic-NESS protocols with the duration or the period $T$ can be written as follows
\be\label{eq:G_q_solution_U_q}
G_{q}(M T) =\la \vec{1}\right|\left(\hat{U}_q(T)\right)^M \left|\vec{p}_{eq}(0)\ra \ .\
\ee
Here, $\left|\vec{p}_{eq}(0)\ra = e^{-\hat e_0}\left|\vec{1}\ra$ is the initial equilibrium probability distribution vector, with $e_{k,nn'} = \delta_{nn'}e_{n}(t_k) = \delta_{nn'}\beta(E_n(t_k)-F(t_k))$.
The evolution operator $\hat{U}_q(t)$ satisfying the same equations~(\ref{eq:rate_eqs_G_q,n}) as $\left|G_q(t)\ra$
is given by the time-ordered exponential
$\hat{U}_q(t)={\rm Texp}(\int_{0}^{t}\hat{\tilde\Gamma}^{(q)}(t)dt)$ and can be written as
a product
\be\label{eq:U_q_prod}
\hat{U}_q(T)= e^{q(\hat e_{K-1}-\hat e_0)}\hat{u}_{K-1} e^{q(\hat e_{K-2}-\hat e_{K-1})}\hat{u}_{K-2}\cdot\ldots\cdot e^{q(\hat e_0-\hat e_1)}\hat{u}_0 \
\ee
compounded of the evolution $e^{q(\hat e_{k}-\hat e_{k+1})}\ne \hat I$ of the generating function of $w_d$ at drive jumps occurring at times $t_k$, $1\leq k\leq K-1$, and of the evolution operators of the probability distribution~(\ref{eq:rate_eqs})
$\hat u_k = \exp[\hat{\Gamma}(t_{k}+0^+) \Delta t_k]$.
Here, we consider discrete time intervals $\Delta t_k = t_{k+1} - t_{k}$, $0=t_0<\ldots<t_{K}=T$, $0\leq k\leq K-1$,
chosen in such a way to neglect variations of $\hat\Gamma$ at each interval $\Delta t_k$, see Fig.~\ref{Fig1_step_drive}(a).
Further, we refer to the drive discretized in such a way as $K$-step drive.
In Eq.~(\ref{eq:G_q_solution_U_q}), the number of periods $M$ equals to unity
for the finite-time protocol, $M=1$, and goes to infinity for periodic-NESS case, $M\to \infty$.

%%%%%%%%%%%%%%%%%%%%%%%%%%%%%%%%%%%%%%%%%%%%%
%%%%%%%%%%%%%%%%%%%%%%%%%%%%%%%%%%%%%

In the periodic-NESS the quantity relevant for fluctuation relations is
the cumulant generating function
\be\label{eq:Delta_q_ln_ep_q}
\Delta_q = \lim_{M\to\infty} \frac{1}{M}\ln G_q(M T) = \ln \ep_q \ ,
\ee
which coincides with the logarithm of the largest eigenvalue $\ep_q$ of the evolution operator $\hat{U}_q$ (see, e.g., \cite{Esposito2013,BaratoChetrite2018}) and independent of the initial
conditions.
In terms of the above mentioned generating functions the integral fluctuation relation (\ref{eq:JE_S_tot}) reads as
\be
\la e^{-w_d}\ra \equiv G_{1}(t) = 1 \Rightarrow \Delta_{1}=0 \ ,
\ee
while the \rev{detailed} ones are
\be\label{eq:Crooks+GC_in_G_q}
G_{q}(t) = G_{1-q}(t) \; {\rm and } \; \Delta_q = \Delta_{1-q} \
\ee
for the finite-time~(\ref{eq:Crooks_W_TRS}) and periodic-NESS~(\ref{eq:GC_W_TRS}) protocols, respectively.

%%%%%%%%%%%%%%%%%%%%%%%%%%%%%%%%%%%%%%%%%%%%%%%%%%%%%%%%%%%%%%%%%%%%%%%%%%%%%%%%%%%%
\section{Time-reversal symmetric drive and beyond}\label{Sec:TRS_and_beyond}
%%%%%%%%%%%%%%%%%%%%%%%%%%%%%%%%%%%%%%%%%%%%%%%%%%%%%%%%%%%%%%%%%%%%%%%%%%%%%%%%%%%%
It is quite obvious that the time-reversal symmetry of the drive is too restrictive for satisfying the \rev{DFRs}
(\ref{eq:Crooks_W_TRS}, \ref{eq:GC_W_TRS}, \ref{eq:Crooks+GC_in_G_q}).
What are more general conditions for which either or both symmetries (\ref{eq:Crooks+GC_in_G_q}) are satisfied?
To answer this non-trivial question, we consider structure of the evolution operator.
Due to the LDB~(\ref{eq:LDB}), the evolution operators at each time step $t_k<t<t_{k+1}$ satisfy the symmetry
\be\label{eq:u_k_symm}
\hat{u}_k = e^{-\hat e_k}\hat u_k^T e^{\hat e_k}
\ee
and the corresponding evolution operator entering the generating function $G_{1-q}(M T) =\la \vec{1}\right|(\hat{U}_{1-q}(T))^M \left|\vec{p}_{eq}(0)\ra$ takes the form after this symmetry transformation
\be\label{eq:U_1-q_prod}
%\hat{U}_q(t)= e^{-\hat e_0/2} e^{(q-1/2)(\hat e_K-\hat e_0)}\hat{u}_q(t_K)e^{(q-1/2)(\hat e_{K-1}-\hat e_K)}\hat{u}_q(t_{K-1})\cdot\ldots\cdot e^{(q-1/2)(\hat e_0-\hat e_1)}\hat{u}_q(t_0)e^{\hat e_0/2} \
\hat{U}_{1-q}(T)= e^{-\hat e_0}e^{q(\hat e_0-\hat e_{K-1})}\hat{u}_{K-1}^T%e^{q(\hat e_{K}-\hat e_{K-1})}\hat{u}^T(t_{K-1})
\cdot\ldots\cdot e^{q(\hat e_1-\hat e_0)}\hat{u}_0^T e^{\hat e_0} \ .
\ee
This leads to the following expressions
for the generating functions in both sides of \rev{DFR~}(\ref{eq:Crooks+GC_in_G_q})
\bea\label{eq:G_q_solution}
G_{q}(T) &=&\la \vec{1}\right|e^{q(\hat e_{K-1}-\hat e_0)}\hat{u}_{K-1} \cdot\ldots\cdot \hat{u}_1 e^{q(\hat e_0-\hat e_1)} \left|\vec{p}_{eq}(0)\ra  \ , \\
\label{eq:G_1-q_solution}
G_{1-q}(T) &=&\la \vec{1}\right|e^{q(\hat e_1-\hat e_0)}\hat{u}_1 \cdot\ldots\cdot \hat{u}_{K-1} e^{q(\hat e_0-\hat e_{K-1})} \left|\vec{p}_{eq}(0)\ra \ .
\eea
One can easily see that the only difference between two expressions is in the inverse order of indices corresponding to the time intervals.

%%%%%%%%%%%%%%%%%%%%%%%%%%%%%%%%%%%%%%%%%%%%%
\begin{figure}[t!]
\hfill\includegraphics[width=0.3\textwidth]{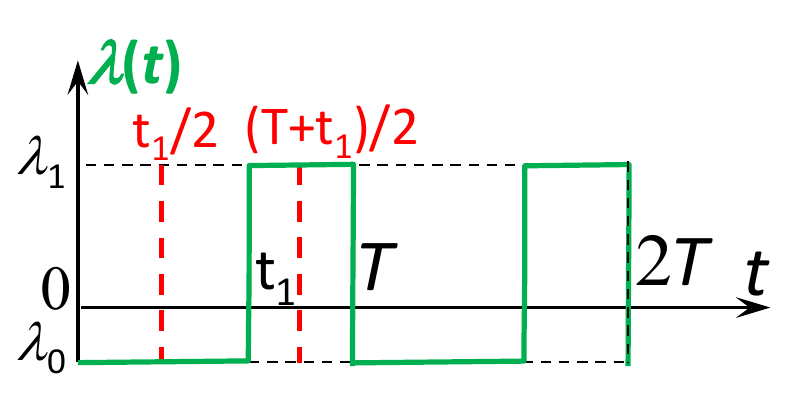}
\caption{Sketch of the two periods of cyclic two-step drive protocol $\lambda(t)$ (green line).
Red vertical lines show positions of time shifts with respect to which the drive is time-reversal symmetric.
} \label{Fig:2step}
\end{figure}
%%%%%%%%%%%%%%%%%%%%%%%%%%%%%%%%%%%%%

%In particular, for the two-step drive, $K=2$, which is always TRS with respect to the \rev{time} shift,
In the particular case of $K=2$ the only evolution operator entering the latter expressions is $\hat u_1$ and $K-1 = 1$, thus the generating functions are (trivially) equal.
Physically in this case of $K=2$, the corresponding two-step drive is TRS with respect to a certain time shift.
Indeed, in this case $\lambda(0<t<t_1) = \lambda_0$ and $\lambda(t_1<t<T)=\lambda_1$ and the time shifts
$t_1/2$ and $(T+t_1)/2$ put the initial time to the middle of one of two plateaux thus making the drive TRS, see Fig.~\ref{Fig:2step}.
As the generating function (\ref{eq:G_q_solution})
does not depend on $\hat u_0$ and, thus, on the zeroth time interval $\Delta t_0$, the symmetry for it is valid in the same way as for the TRS drive protocol.
%due to the independence of $G_q(T)$ of the duration of the zeroth time interval $\Delta t_0$.
It may be confusing why $G_q(T)$ is independent of the zeroth time interval $\Delta t_0$, but explicitly depends on the last one $\Delta t_{K-1}$.
The answer to this question is hidden in the choice of the time discretization.
Indeed, we have chosen the discretized protocol to start with the plateau followed by the instantaneous jump at $t_{k+1}$ at each time interval $t_k<t\leq t_{k+1}$, Fig.~\ref{Fig1_step_drive}(a). As the system is in equilibrium initially for the finite-time protocol, $M=1$, the absence of the drive in $0<t<t_1$ changes nothing.
In an alternative discretization shown in Fig.~\ref{Fig1_step_drive}(b), when the jumps in drive are followed by plateaux,
$G_q(T)$ depends explicitly on $\Delta t_0$, but not on $\Delta t_{K-1}$ as the relaxation at $t_{K-1}<t<T$ does not affect the dissipated work.
In general continuous drive both possible plateaux in the beginning and in the end of the drive do not affect dissipated work as the control parameter is constant, Fig.~\ref{Fig1_step_drive}(c).
Here and further, we stick to the first variant of discretization shown in Fig.~\ref{Fig1_step_drive}(a) for clarity.

For the general TRS drive all time intervals are coupled in pairs $\hat{u}_k =\hat{u}_{K-k}$, $\hat e_{K-k} = \hat e_{k}$
and, thus, the expressions (\ref{eq:G_q_solution}, \ref{eq:G_1-q_solution}) are equal
and \rev{finite-time DFR} in (\ref{eq:Crooks+GC_in_G_q}) is obviously satisfied.
The corresponding evolution operators $\hat{U}_{q}(T)$ and $\hat{U}_{1-q}(T)$ simply relate to each other
\be\label{eq:U_1-q_sym}
\left(\hat{U}_{1-q}^T\right)^M(T) = \hat C \left(\hat{U}_{q}(T)\right)^M\hat C^{-1} \ ,%\hat{u}_0^{-1}e^{-\hat e_0} \ .
\ee
with $\hat C = e^{\hat e_0} \hat{u}_0$ for any $M$.
Thus, \rev{asymptotic DFR} in (\ref{eq:Crooks+GC_in_G_q}) is also satisfied as both the initial conditions and the evolution $\hat C$
give only subleading contributions to $\Delta_q$ in the limit $t\to \infty$.

This \rev{asymptotic DFR} in (\ref{eq:Crooks+GC_in_G_q}) is also preserved in more general case, when the relation~(\ref{eq:U_1-q_sym}) between evolution operators $\hat{U}_{1-q}(T)$ and $\hat{U}_{q}(T)$ holds with an arbitrary matrix $\hat C$ which depends on $q$ and on the protocol at one period, but not on the number of periods $M$.
If on top of that we
initialize the system in such a way that the vectors
$\left|\vec{p}_{eq}(0)\ra\equiv e^{-\hat e_0}\left|\vec{1}\ra$ and $\la\vec{1}\right|$
are the right and left eigenvectors of $\hat C$, respectively, with the same eigenvalue $c$
\be\label{eq:C-symm_init_cond}
\hat C \left|\vec{p}_{eq}(0)\ra = c \left|\vec{p}_{eq}(0)\ra, \quad \la\vec{1}\right|\hat C = \la\vec{1}\right|c \ ,
\ee
the expressions (\ref{eq:G_q_solution}, \ref{eq:G_1-q_solution}) become equal.

From this perspective one might come to a quite natural conclusion that the symmetry of the cumulative function $\Delta_q = \Delta_{1-q}$ in periodic-NESS protocols is less restrictive than the one of the generating function $G_q = G_{1-q}$ in finite-time protocols
as the former does not have any conditions on the initial distribution (cf. the discussion of the role of initial conditions in NESS~\cite{VerleyLacoste2012}).
However, in general it is not so clear.
Indeed, the condition~(\ref{eq:Crooks+GC_in_G_q}) for $M>1$ crucially depends on $\Delta t_0$ via the step evolution operator $\hat u_0$, while expressions (\ref{eq:G_q_solution}, \ref{eq:G_1-q_solution}) do not.
To clarify this statement, we derive a general relation between the generating function $G_q(T)$ and the trace of the evolution operator for $\Delta t_0\to \infty$
\be\label{eq:G_q_tr_U_q}
G_{q}(T) \equiv\la \vec{1}\right|\hat{U}_q(T) \left|\vec{p}_{eq}(0)\ra = \lim_{\Delta t_0\to \infty} \tr \hat{U}_q(T) \ .
\ee
This is the main result of our paper, which works for any classical Markovian $N$-level system obeying rate equations (\ref{eq:rate_eqs}).

%%%%%%%%%%%%%%%%%%%%%%%%%%%%%%%%%%%%%%
\begin{table*}[t]
 \begin{tabular}{ l | c | c | c | c }
 & Crooks
 & TRS Crooks
 & $G_q$ symmetry
 & $G_q$ expression
 \\
 \hline			
Finite-time drive & $\frac{P(W)}{\bar P(-W)} = e^{\beta (W-\Delta F)}$ & $\frac{P(W)}{P(-W)} = e^{\beta (W-\Delta F)}$ & $G_q = G_{1-q}$ &
%$G_q = \lim_{\Delta t_0\to \infty} \tr \hat{U}_q$
$G_q \stackrel{\Delta t_0\to\infty}{\longleftarrow} \tr \hat{U}_q$\footnote{Original result derived in this paper}
\\
Periodic NESS &  $\frac{1}{t} \ln \frac{P(wt)}{\bar P(-wt)} \stackrel{t\to\infty}{\longrightarrow} \beta w$ &  $\frac{1}{t} \ln \frac{P(wt)}{P(-wt)} \stackrel{t\to\infty}{\longrightarrow} \beta w$ & $\Delta_q = \Delta_{1-q}$ & $\Delta_q = \max {\rm spec}[\hat U_q]$
\\
 \hline
 \end{tabular}
 \caption{Summary of finite-time and periodic NESS fluctuation theorems. The notation "max spec" means the maximal eigenvalue in the spectrum of an operator.}
 \label{Table:Crooks_summary}
\end{table*}
%%%%%%%%%%%%%%%%%%%%%%%

The origin of this relation lies in the structure of rate equations with constant tunneling rates $\Gamma_{nn'}$, for example, at a certain step $t_k<t<t_{k+1}$.
Indeed, the eigenvalues $\gamma_m(t_k)\leq0$ of the rate matrix $\hat\Gamma(t_k+0)$ are negative, except one single zero value $\gamma_0 = 0$ corresponding to
the unit left eigenvector $\la \vec{1}\right|$ and to the instantaneous equilibrium distribution vector $\left|\vec{p}_{eq}(t_k)\ra$ as a right eigenvector.
Thus, the evolution operator reads
\be\label{eq:step_u(t)}
\hat u_k(\Delta t_k) = \left|\vec{p}_{eq}(t_k)\ra\la \vec{1}\right| + e^{-|\gamma_{\min}|\Delta t_k}\hat{\delta u}_k(\Delta t_k) \ ,
\ee
where $\gamma_{\min}(t_k) = \rev{\max}_{m\ne 0} \gamma_m(t_k) <0$
and $\hat{\delta u}_k(\Delta t_k)$ is the matrix with non-increasing matrix elements.
The second term in~(\ref{eq:step_u(t)}) decays exponentially fast to zero with increasing $\Delta t_0$.
Thus, considering the limit $\Delta t_0\to\infty$ in r.h.s. of~(\ref{eq:G_q_tr_U_q}) and substituting expressions (\ref{eq:G_q_solution}) and (\ref{eq:U_q_prod}) in l.h.s. and r.h.s., one can easily prove the relation~(\ref{eq:G_q_tr_U_q}).

From Eq.~(\ref{eq:G_q_tr_U_q}) one can conclude that the finite-time fluctuation relation~(\ref{eq:Crooks_W_TRS}) is satisfied
as soon as the trace of the evolution operator in the limit $\Delta t_0\to \infty$ \rev{satisfies the symmetry}
\be
\lim_{\Delta t_0\to \infty} \tr \hat{U}_q(T) = \lim_{\Delta t_0\to \infty} \tr \hat{U}_{1-q}(T) \ .
\ee
On the other hand, the validity of the asymptotic fluctuation relation~(\ref{eq:GC_W_TRS}) depends
not only on the evolution operator trace symmetry, but on the symmetry of its maximal eigenvalue~(\ref{eq:Delta_q_ln_ep_q}).
This shows that neither of \rev{DFRs} in~(\ref{eq:Crooks+GC_in_G_q}) implies the other.
The general results on the detailed fluctuation theorems for the dissipated work known in the literature or derived in this section
are summed up in Table~\ref{Table:Crooks_summary}.

A particular case of the symmetry~(\ref{eq:U_1-q_sym})
relating the step evolution operators $\hat u_k$ and $\hat u_{K-k}$
and generalizing the TRS drives is considered in~\ref{App_Sec:B_q_sym}.
This case provides an example when the asymptotic fluctuation theorem implies finite-time counterpart, unlike the results in NESS~\cite{VerleyLacoste2012}.

As shown in the next section, the satisfaction of the finite-time fluctuation theorem does not imply the same in the asymptotic long-time limit
even in the simplest possible example of a classical Markovian two-level system.

%%%%%%%%%%%%%%%%%%%%%%%%%%%%%%%%%%%%%%%%%%%%%%%%%%%%%%%%%%%%%%%%%%%%%%%%%%%%%%%%%%%%
\section{Two-level system}\label{Sec:TLS}
%%%%%%%%%%%%%%%%%%%%%%%%%%%%%%%%%%%%%%%%%%%%%%%%%%%%%%%%%%%%%%%%%%%%%%%%%%%%%%%%%%%%
Two-level systems are special in several aspects.
First, any rate matrix $\hat \Gamma$ in two-level systems satisfies LDB condition with certain energy difference $\beta(E_1 - E_0)\equiv \ln \left[\Gamma_{01}/\Gamma_{10}\right]$ normalized to temperature. Moreover, any probability distribution can be considered as thermal with a certain parameter $\beta(E_1 - E_0)$ possibly different from the above one~\footnote{As one of consequences, in two-level systems it is possible to write fluctuation relations not only for thermodynamic quantities, but even for the finite-time average of the charge state~\cite{Singh2016}}.
In both cases the energy difference $E_1 - E_0$ might be not equal to the physical energy difference in non-equilibrium conditions, but as the two-level system has the only control parameter $2\lambda=\beta(E_1 - E_0)\equiv \ln \left[\Gamma_{01}/\Gamma_{10}\right]$ we will use it further.
Second, there are only two drive symmetries of the kind of (\ref{eq:U_1-q_sym}), TRS $\lambda(T-t)=\lambda(t)$ and anti-TRS drive $\lambda(T-t)=-\lambda(t)$.
The difference between symmetric and anti-symmetric drives is subtle as the exchange of energies keeps the overall spectrum intact.
However, one should take into account that the non-adiabatic exchange of energy levels affects the occupation probabilities $p_n(t)$.
For example, if one prepares a two-level system in equilibrium with a certain ground $E_0$ and excited $E_1$ state energies and then suddenly exchange them ($\lambda(T/2-0) = -\lambda(T/2+0)$), the system would not be in the same equilibrium state and will decay to the new equilibrium after such quench perturbation.

%%%%%%%%%%%%%%%%%%%%%%%%%%%%%%%%%%%%%%%%%%%%%
\begin{figure}[t!]
\hfill\includegraphics[width=\textwidth]{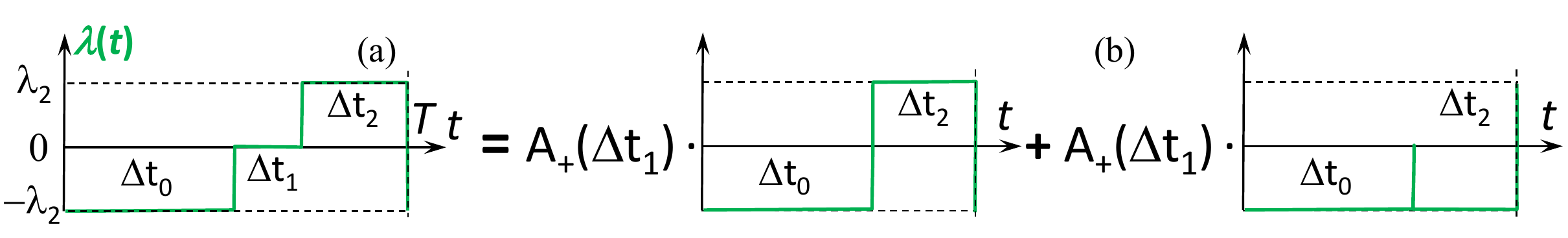}
\caption{(a) Sketch of the cyclic three-step drive protocol $\lambda(t)$ (green line) and (b) its decomposition~(\ref{eq:G_q_solution_3-step})
into two-step cyclic and one-step non-cyclic drives \rev{with the coefficients $A_\pm(\Delta t_1)$ of the expansion (\ref{eq:step_u0(t)}) of the evolution operator $\hat u_1(\Delta t_1)$}.
} \label{Fig:3step}
\end{figure}
%%%%%%%%%%%%%%%%%%%%%%%%%%%%%%%%%%%%%

To start with in this section we first go beyond symmetric and anti-symmetric drives mentioned above.
As shown in the previous section any two-step drive, $K=2$, is TRS and thus it leads to \rev{DFR} (\ref{eq:Crooks+GC_in_G_q}) without any additional conditions (see, e.g.,~\cite{Esposito2013}).
Therefore, we do one step beyond and provide an example
of the simplest non-TRS drive, namely, three-step drive, $K=3$, Fig.~\ref{Fig:3step}(a), and consider general conditions under which
this drive satisfies both relations (\ref{eq:Crooks_W_TRS}) and (\ref{eq:GC_W_TRS}).

As follows from the calculations given in \ref{App_Sec:3-step},
the necessary and sufficient condition for both \rev{DFRs} (\ref{eq:Crooks+GC_in_G_q}) restrict the values of $\lambda(t)$ at the drive steps to be the following up to any permutation between steps
\be\label{eq:3step_cond}
\lambda_0 = -\lambda_2, \quad \lambda_1 = 0 \ .
\ee
The surprising thing here is that the above condition is independent not only of the zeroth time interval, but of {\it all} the time durations.
One could understand this fact if for the generating function symmetry $G_q = G_{1-q}$ one needed to begin driving from
the degeneracy point $\lambda=0$ when both energies are equal in order to have nearly anti-TRS drive. However, for this, one has to make
two other time intervals to be equal, which is not the case.
Even more surprising thing is that the symmetry $G_q = G_{1-q}$ is valid for any permutations and time shifts of the drive.

The origin of this emerging symmetry is hidden in the structure of the evolution operator $\hat u_1(\Delta t_1)$ at $\lambda = 0$.
Indeed, due to the equal values of both incoming rates $\Gamma_{01} = \Gamma_{10}\equiv |\gamma_{\min}|/2$ this evolution operator can be expanded into
the superposition of the unity matrix $\hat I$ and the \rev{Pauli} matrix $\sigma_x$ reordering the energy levels $E_n$ in the inverse order
\be\label{eq:step_u0(t)}
\hat u_1(\Delta t_1) = A_+ \hat I + A_-\hat \sigma_x \ ,
\ee
with $2A_\pm = 1\pm e^{-|\gamma_{\min}|\Delta t_1}$.
As a result, the generating function~(\ref{eq:G_q_solution}) splits into the sum of two-step cyclic and one-step acyclic drives corresponding to the first and second terms in r.h.s. of both following expressions, respectively (see Fig.~\ref{Fig:3step} for details)
\bea\label{eq:G_q_solution_3-step}
G_{q}(T) =\la \vec{1}\right|e^{-2q\hat e_{0}}\hat{u}_{2} e^{q\hat e_0} (A_+\hat I+A_-\sigma_x) e^{q\hat e_0} \left|\vec{p}_{eq}(0)\ra \nonumber \\
=A_+\la \vec{1}\right|e^{-2q\hat e_{0}}\hat{u}_{2} e^{2q\hat e_0} \left|\vec{p}_{eq}(0)\ra
+A_-\la \vec{1}\right|e^{2q\hat e_{0}} \left|\vec{p}_{eq}(0)\ra
\ , \\
G_{1-q}(T) =\la \vec{1}\right|e^{-q\hat e_0} (A_+\hat I+A_-\sigma_x) e^{-q\hat e_0}\hat{u}_{2}e^{2q\hat e_{0}}\left|\vec{p}_{eq}(0)\ra \nonumber \\
=A_+\la \vec{1}\right|e^{-2q\hat e_0}\hat{u}_{2}e^{2q\hat e_{0}} \left|\vec{p}_{eq}(0)\ra
+A_-\la \vec{1}\right|e^{2q\hat e_{0}}\left|\vec{p}_{eq}(0)\ra
\ .
\eea
This example opens the way to form non-TRS drives satisfying TRS versions~(\ref{eq:Crooks_W_TRS}, \ref{eq:GC_W_TRS}) of fluctuation theorems
and motivates the studies in such simple models the first-passage time distribution~\cite{Singh_1st_passage,Singh_Thesis} and the analogy with multifractality~\cite{MF_analogy} mentioned in the introduction.

Moreover, two-level systems allow one to find an explicit relation between the \rev{finite-time and asymptotic DFR}~(\ref{eq:Crooks+GC_in_G_q}).
Indeed, we show below that if the asymptotic fluctuation theorem is valid $\Delta_q = \Delta_{1-q}$ for {\it two} drive protocols which differ only in the duration $\Delta t_0$ and $\Delta t_0'$ of the zeroth time interval,
then its finite-time counterpart $G_{q} = G_{1-q}$ is also valid.
Moreover in this case the asymptotic fluctuation theorem is valid $\Delta_q = \Delta_{1-q}$ for {\it any} $\Delta t_0$.

Surprisingly this statement also works in another direction:
if asymptotic $\Delta_q = \Delta_{1-q}$ and finite-time $G_{q} = G_{1-q}$ fluctuation theorems are satisfied for a certain drive protocol,
then they are valid for such protocol with {\it any} $\Delta t_0$.

The origin of this relations contains several ingredients.
First one is the expression for the generating function $G_q$, Eq.~(\ref{eq:G_q_solution_U_q}), through the trace of the evolution operator
(\ref{eq:G_q_tr_U_q}), see the derivation in the previous section.

The second ingredient is that for a two-level system the validity of the symmetry $\Delta_q = \Delta_{1-q}$
is solely governed by the validity of the symmetry for the trace of the evolution operator
\be\label{eq:tr_U_q_symm}
\tr \hat U_q(T) = \tr \hat U_{1-q}(T) \ .
\ee
Indeed, for any protocol and any classical Markovian $N$-level system the determinant of the evolution operator is $q$-independent
and given by
$\det\hat{U}_q(t) = e^{-\int_0^t \tr\hat\Gamma(t')dt'}$ (see~\ref{App_Sec:gen_func} or~\cite{Esposito2013} for details).
In the two-level system the eigenvalues of $\hat U_q(T)$ are determined only by $\det\hat{U}_q(t)$ and $\tr \hat U_q(T)$,
$2\ep_q = \tr \hat U_q(T) + \sqrt{ \left [ \tr \hat U_q(T)\right ]^{2} - 4 \det \hat U_q(T) }$, thus $\Delta_q$ as the maximal eigenvalue among two is symmetric, $\Delta_q = \Delta_{1-q}$, if and only if Eq.~(\ref{eq:tr_U_q_symm}) is satisfied.

The third and final ingredient for this calculation is the expression (\ref{eq:step_u(t)}) for the step evolution operator similar to (\ref{eq:step_u0(t)}),
where $-\gamma_{\min} = \Gamma_{01}(t_k)+\Gamma_{10}(t_k)>0$ and
$\hat{\delta u}_k$ is the constant matrix for two-level systems.

Combining ingredients~(\ref{eq:G_q_tr_U_q}, \ref{eq:step_u(t)}, \ref{eq:tr_U_q_symm}) together one can express $G_q(T)$ via
two $\tr \hat U_q(T, \Delta t_0)$ with different values $\Delta t_0$ and $\Delta t_0'$ of the zeroth time interval duration as follows
\be\label{eq:G_q_2_tr_U_q}
G_q(T) = \frac{\tr \hat U_q(T, \Delta t_0)-e^{-|\gamma_{\min}(0)|(\Delta t_0-\Delta t_0')}\tr \hat U_q(T, \Delta t_0')}
{1-e^{-|\gamma_{\min}(0)|(\Delta t_0-\Delta t_0')}} \ .
\ee
Analogously one can express $\tr \hat U_q(T, \Delta t_0'')$ through the same functions, see~\ref{App_Sec:G_q_trU_q}.

However, such analysis cannot be repeated for a general classical Markovian $N$-level system.
Indeed, as shown in the previous section the expression~(\ref{eq:G_q_tr_U_q}) is valid,
while (\ref{eq:tr_U_q_symm}) is only necessary, but not sufficient condition for $\Delta_q = \Delta_{1-q}$
as not only determinant and trace govern the maximal eigenvalue of the matrix $\hat U_q(T)$.
The expression (\ref{eq:G_q_2_tr_U_q}) also cannot be written as, in general, the matrix structure of $\hat{\delta u}_k$ is time-dependent.

The only thing which one can derive is that the sufficient condition to have $G_q = G_{1-q}$
is the presence of the symmetry $\Delta_q = \Delta_{1-q}$ for $N$ different zeroth time interval durations (leading to (\ref{eq:tr_U_q_symm}) for each of them). This sufficient condition comes from the fact that the matrix $\hat{\delta u}_k$ can be written as the sum of constant matrices with $N-1$ different exponentially decaying prefactors and, thus, one can derive expression for $G_q$ analogous to (\ref{eq:G_q_2_tr_U_q}), but it will include $N$ traces
$\tr\hat U_q(T)$ for the protocols with different $\Delta t_0$ in order to remove all $N-1$ exponentially decaying components of $\hat{\delta u}_k$.

This provides a hint that the symmetries both in finite-time fluctuation relations and in their periodic-NESS counterparts
become more restrictive with increasing system degrees of freedom, but it cannot completely resolve the question about the relation between them.

%%%%%%%%%%%%%%%%%%%%%%%%%%%%%%%%%%%%%%%%%%%%%%%%%%%%%%%%%%%%%%%%%%%%%%%%%%%%%%%%%%%%
\section{Conclusion}\label{Sec:Conclusion}
%%%%%%%%%%%%%%%%%%%%%%%%%%%%%%%%%%%%%%%%%%%%%%%%%%%%%%%%%%%%%%%%%%%%%%%%%%%%%%%%%%%%
To sum up, in this paper, the relations between finite-time~(\ref{eq:Crooks_W_TRS}) and infinite-time~(\ref{eq:GC_W_TRS}) fluctuation relations are considered.
We are motivated to focus on the versions of these fluctuation theorems coinciding in their form with the ones for time-reversal symmetric drives as they provide the solid ground both for the straightforward calculations of first-passage-time distribution~\cite{Singh_Thesis,Singh_1st_passage} and
for the unexpected analogy of the work statistics with the multifractality of the wavefunctions close to the Anderson localization
transition~\cite{MF_analogy}.

In the general case of a classical Markovian $N$-level system, we derive the condition~(\ref{eq:U_1-q_sym})
with an arbitrary matrix $\hat C$ depending on $q$ and on the protocol at one period, but not on the number of periods $M$
to satisfy an infinite-time fluctuation theorem~(\ref{eq:GC_W_TRS}).
Also we provide the sufficient condition (\ref{eq:C-symm_init_cond}) for the corresponding finite-time fluctuation theorem~(\ref{eq:Crooks_W_TRS})
posing additional restrictions on the initial distribution similarly to~\cite{VerleyLacoste2012}.
On the other hand, the particular case~(\ref{eq:B_q-symm}, \ref{eq:B_q-symm_init_cond}) of the above mentioned symmetries is considered in~\ref{App_Sec:B_q_sym} and provides an example when the finite-time fluctuation theorem is less restrictive than its asymptotic counterpart.

In the particular case of a two-level system the explicit relation~(\ref{eq:G_q_2_tr_U_q})
between finite-time~(\ref{eq:Crooks_W_TRS}) and infinite-time~(\ref{eq:GC_W_TRS}) fluctuation relations is found.
Its formulation reads in two ways:
If the asymptotic fluctuation theorem is valid $\Delta_q = \Delta_{1-q}$ for
{\it two} drive protocols which differ only in the duration $\Delta t_0$ and $\Delta t_0'$ of the zeroth time interval,
then its finite-time counterpart $G_{q} = G_{1-q}$ is valid as well as the asymptotic one for such protocol with {\it any} $\Delta t_0$.
If asymptotic $\Delta_q = \Delta_{1-q}$ and finite-time $G_{q} = G_{1-q}$ fluctuation theorems are satisfied for a certain drive protocol,
then they are valid for such protocol with {\it any} $\Delta t_0$.
Additionally, the class of drive protocols satisfying the above mentioned relations is extended from the time-reversal-(anti)symmetric ones and
an example of the simplest non-time-reversal-(anti)symmetric drive is given.

%%%%%%%%%%%%%%%%%%%%%%%%%%%%%%%%%%%%%%%%%%%%%%%%%%%%%%%%%%%%%%%%%%%%%%%%%%%%%%%%%%%%
\section{Acknowledgements}
%%%%%%%%%%%%%%%%%%%%%%%%%%%%%%%%%%%%%%%%%%%%%%%%%%%%%%%%%%%%%%%%%%%%%%%%%%%%%%%%%%%%
We are grateful to {\'E}. Rold{\'a}n and J. P. Pekola for stimulating discussions.
I.~M.~K. acknowledges the support of German Research Foundation (DFG) Grant No. KH~425/1-1 and the Russian Foundation for Basic Research Grant No.~17-52-12044.
In the part concerning two-level system directly related to the Coulomb-blockaded devices, the work was supported by Russian Science Foundation under Grant No. 17-12-01383.
%the Russian Science Foundation, Grant No. 17-12-01383 (I.M.K.).

\appendix

\section{Rate equations and generating functions}\label{App_Sec:gen_func}
In this Appendix section, we give detailed calculations of the probability distribution functions $P(X)$ of a certain stochastic quantity $X$ and
of the corresponding generating function $G_q$ based on the rate equations~(\ref{eq:rate_eqs}).
As in the case of the dissipated work the probability distribution $P(X)$ itself does not determine explicitly
the system state $n(t)$ one have to generalize it to the $n$-resolved distribution function
$\left|\vec{P}(X)\ra = (P_0(X),\ldots,P_{N-1}(X))$, with the components defined as
\be
P_n(X=x) = \langle \delta(X-x)\delta_{n,n(t)}\rangle\ .
\ee
The distribution function is given by the sum $P(X) = \la \vec{1}|\vec{P}(X)\ra\equiv \sum_n P_n(X)$.

In the special case of the work $X=W$,
one can write rate equations for $P_n(W)$ explicitly~\cite{Esposito2013}
\be\label{eq:rate_eqs_P_n(W)}
\frac{d}{dt} \left|\vec{P}(W,t)\ra = \hat\Gamma(t)\left|\vec{P}(W,t)\ra-\frac{\partial}{\partial W}\left[\hat{\dot{W}} \left|\vec{P}(W,t)\ra\right]
\ee
as the work rate in the certain state $\dot W_n\equiv \left.\frac{dW}{dt}\right|_{n(t)=n} = \frac{\partial E_n}{\partial \lm}\frac{d\lm}{dt}$
(written in the matrix form $\dot W_{n,n'}\equiv\delta_{n,n'}\dot W_n$) is a deterministic function of the system state $n(t)$.
As work performed on the system at time $t=0$ is zero the initial condition for $\left|\vec{P}(W,t)\ra$ reads as
$\left|\vec{P}(W,0)\ra=\delta(W)\left|\vec{p}(0)\ra$.
This analysis also works for any quantity $X$ with the same property of $\dot X_n$.

In general, it is impossible to write the rate equation for $\left|\vec{P}(X)\ra$ itself, but one can do it
for the $n$-resolved generating function of the variable $X$ defined as the Laplace transform of the latter
\be\label{eq:G^X_q_vec}
\left|\vec{G}_{q}\ra = \int \left|\vec{P}(X)\ra e^{-q X} dX \ , \quad G_{q,n} = \langle e^{-q X(t)}\delta_{n,n(t)}\rangle \ .
\ee

Indeed, considering the system state trajectory $\{n(t)\}$ as a set of jumps from $n_{j-1}$ to $n_j$ occurred at time instants $t_j^{(J)}$, $j=1,N_J$, $t_j^{(J)}<t_{j+1}^{(J)}$, $t_0^{(J)} = 0$, $t_{N_J+1}^{(J)} = t$, Fig.~\ref{Fig0_lambda_n_W_Q},
one can write
the trajectory probability measure
explicitly
\bea\label{eq:P_M-jump}
P_{N_J} (t; n_0, t_0^{(J)}, n_1, t_1^{(J)},\ldots, n_M, t_M^{(J)}) &=& p_{n_0}(t_0^{(J)})e^{-\int_{t_0^{(J)}}^{t_1^{(J)}}\Gamma_{n_0,n_0}(t')dt'}\times \nonumber \\
&\prod_{j=1}^{N_J}&\Gamma_{n_j,n_{j-1}}(t_j^{(J)}) e^{-\int_{t_j^{(J)}}^{t_{j+1}^{(J)}}\Gamma_{n_j,n_j}(t')dt'} \ ,
\eea
which is the product of the
probabilities $\exp\left[-\int_{t_j^{(J)}}^{t_{j+1}^{(J)}}\Gamma_{n_j,n_j}(t')dt'\right]$ to have {\it no jumps} in the system in the time interval $(t_j^{(J)}, t_{j+1}^{(J)})$ provided the system was in the state $n_j$ at time instant $t_j^{(J)}$
and the conditional probabilities $\Gamma_{n_j,n_{j-1}}(t_j^{(J)})dt_j^{(J)}$ to have a jump from $n_{j-1}$ to $n_{j}$ in the time interval $(t_{j}^{(J)},t_{j}^{(J)}+dt_{j}^{(J)})$
provided there was no jumps in the interval $(t_{j-1}^{(J)}, t_{j}^{(J)})$.
As a result, the rate equations~(\ref{eq:rate_eqs}) can be easily derived from this expression
with help of averaging over $P_{N_J}$ of the definition of the probability distribution $p_n(t) = \la\delta_{n,n(t)}\ra$, see, e.g.,~\cite{Bagrets_Nazarov2003}.

To write the rate equation for the $n$-resolved generating function $G_{q,n}=\la e^{-q X(t)}\delta_{n,n(t)}\ra$
of the piecewise deterministic stochastic process~\cite{Breuer_Petruccione_book}
$X(t)$, one should average $e^{-q X(t)}\delta_{n,n(t)}$ over the same distribution~(\ref{eq:P_M-jump}).
For this, one needs to write the expression for $X(t)$ at the same state trajectory
\be\label{eq:X(t)}
X(t) = \sum_j \left[\Delta X_{n_{j-1}\to n_j}(t_j^{(J)})+\int_{t_{j-1}^{(J)}}^{t_j^{(J)}} \dot X_{n_{j-1}}(t') dt'\right] \ .
\ee
Equation~(\ref{eq:X(t)}) has both the deterministic contributions $\dot X_{n}(t)$ at fixed $n$ and the stochastic jumps $\Delta X_{n\to n'}(t)$ due to the jumps in $n(t)$
(like for the total entropy production $\Delta s_{tot}$).
These contributions enter the generating function expression just by modifying the rates
\be\label{eq:Gamma^X_n->n'_X}
\Gamma^{(q)}_{n,n'}(t) = \Gamma_{n,n'}(t)e^{-q \Delta X_{n'\to n}(t)} \ , \quad
\Gamma^{(q)}_{n,n}(t) = \Gamma_{n,n}(t) + q  \dot X_n(t) \ .
\ee

Thus, with use of the standard trajectory representation of the Markov jump processes which is widely used in the full counting statistics (see,
e.g.,~\cite{Bagrets_Nazarov2003}),
we derive the rate equations~(\ref{eq:rate_eqs_G_q,n} for the generating function in the form of (\ref{eq:rate_eqs})
\be\label{eq:rate_eqs_G^X_q,n}
\frac{d}{dt}\left|\vec{G}_{q}(t)\ra = \hat\Gamma^{(q)}(t)\left|\vec{G}_{q}(t)\ra \ ,
\ee
with the modified rates (\ref{eq:Gamma^X_n->n'_X}) and the initial condition $\left|\vec{G}_{q}(0)\ra=\left|\vec{p}(0)\ra$ provided
$X(0)=0$. Note that unlike Eq.~(\ref{eq:rate_eqs}) the latter equation does not conserve normalization condition as $\Gamma^{(q)}_{n,n} \ne
\sum_{n'\ne n} \Gamma^{(q)}_{n',n}$.

For the quantities which depend only on the change of the system state $n(t)$
(like the heat $Q$ or the environment entropy production $\Delta s_m$),
only the incoming rates are modified by the exponential factor depending on the size of the corresponding jump $\Delta X_{n'\to n}(t)$
\be\label{eq:Gamma^X_n->n'_Q-like}
\Gamma^{(q)}_{n,n'}(t) = \Gamma_{n,n'}(t)e^{-q \Delta X_{n'\to n}(t)} \ . %\ , \quad \Gamma^{X,q}_{n} = \Gamma_{n}\ .
\ee
Unlike this, for the quantities (like the work $W$) for which rate $\dot X_n(t)$ is a deterministic function of $n(t)$ only the escape rates should be
modified
\be\label{eq:Gamma^X_n->n'_W-like}
\Gamma^{(q)}_{n,n}(t) = \Gamma_{n,n}(t) + q  \dot X_n(t) \ .
\ee

The probability distribution of $X$
\be\label{eq:P(x)}
P(X) = \frac{1}{2\pi i}\lim_{Q\to\infty}\int_{\chi-iQ}^{\chi+iQ} G_{q}(t) e^{q X} d q \ .
\ee
is given by the inverse Laplace transform of the generating function, where $\chi$ is greater than the real part of all singularities of
$G_{q}(t)$ as a function of $q$ and
\be\label{eq:G^X_q}
G_{q}(t) =\la \vec{1}|\vec{G}_q(t)\ra\equiv \sum_n G_{q,n}(t) \ .
\ee

Note that the deterministic part $X_n(t) = \int^t \dot X_{n}(t)$ of the piecewise deterministic stochastic process (\ref{eq:X(t)})
can be absorbed by the following transformation
\be
%\tilde G_{q,n}(t) = G_{q,n}(t)e^{q X_n(t)} \ ,
\left|\tilde{\vec{G}}_q(t)\ra = e^{q \hat X(t)}\left|\vec{G}_q(t)\ra \ .
\ee
restoring a simple jump process with the jump size being the sum of two contributions $\Delta X_{n'\to n}(t)+(X_{n'}(t)-X_n(t))$.
Here, $X_{n,n'}(t) \equiv \delta_{n,n'}X_n(t)$
and the l.h.s. satisfies the rate equations (\ref{eq:rate_eqs_G^X_q,n}) with the rates replaced by
\be\label{eq:tilde_Gamma_rates}
\tilde \Gamma^{(q)}_n = \Gamma_n \; {\rm and} \; \tilde \Gamma^{(q)}_{n,n'} = \Gamma_{n,n'}e^{-q\left[\Delta X_{n'\to
n}(t)+X_{n'}(t)-X_n(t)\right]} \ .
\ee
The price paid for this simplification is the modification of the initial conditions
\be
%\tilde G_{q,n}(0) = p_n(0) e^{q X_n(0)} \ .
\left|\tilde{\vec{G}}_q(0)\ra = e^{q \hat X(0)}\left|\vec{p}(0)\ra \ .
\ee

The evolution operator $\hat{U}_q(t)$ entering the expression~(\ref{eq:G_q_solution_U_q}) for the generating function $G_{q}(M T)$
satisfies the same rate equations~(\ref{eq:rate_eqs_G_q,n}, \ref{eq:rate_eqs_G^X_q,n}) as $\left|G_q(t)\ra$.
Thus, the measure of phase volume contraction of the system stochastic dynamics, namely,
the determinant of the evolution operator $\det\hat{U}_q(t)$ satisfies the following rate equation
\be\label{eq:rate_eqs_det_U^X_q}
\frac{d}{dt}\det\hat{U}_q(t) = \tr\hat{\Gamma}^{(q)}(t)\det\hat{U}_q(t)
\ee
and does not depend on $q$ as $\tr\hat{\Gamma}^{(q)}(t) =\tr\hat\Gamma(t) = \sum_{n} \Gamma_{n,n}(t)$
\be\label{eq:det_U_q}
\det\hat{U}_q(t) = e^{-\int_0^t \tr\hat\Gamma(t')dt'}\equiv e^{-\tau(t)}\leq 1 \ .
\ee
The function $\tau(t)$ gives a certain rescaled ``time'' (analogous to the entropic time in Ref.~\cite{Pigolotti_PRL}),
which sets time of the fastest decay to unity. Note that the time-reversal transformation changing $t$ by $t_{\max}-t$
changes the rescaled time $\tau(t)$ by $\tau(t_{\max})-\tau(t)$ as well.

\section{Example of symmetry~(\ref{eq:U_1-q_sym})}\label{App_Sec:B_q_sym}
The particular example of the symmetry~(\ref{eq:U_1-q_sym}) for the asymptotic fluctuation theorem~(\ref{eq:GC_W_TRS})
relating the step evolution operators $\hat u_k$ and $\hat u_{K-k}$
in time intervals $\Delta t_k$ and $\Delta t_{K-k}$
and generalizing the TRS drives can be written as follows
\be\label{eq:B_q-symm}
e^{q\hat e_k}\hat{u}_k e^{-q\hat e_k} = \hat B_q e^{q\hat e_{K-k}}\hat{u}_{K-k} e^{-q\hat e_{K-k}} \hat B_q^{-1} \ .
\ee
Here, $\hat B_q$ is a certain time-independent matrix.
This symmetry corresponds to the following expression for the matrix
$\hat C = e^{\hat e_0} \hat{u}_0 e^{-q \hat e_0}\hat B_q e^{q \hat e_0}$ from~(\ref{eq:U_1-q_sym}) if the matrix $e^{-q \hat e_0}\hat B_q e^{q \hat e_0}$ commute with $\hat{u}_0$
\be\label{eq:B_q_u0_commute}
e^{-q \hat e_0}\hat B_q e^{q \hat e_0}\hat{u}_0=\hat{u}_0e^{-q \hat e_0}\hat B_q e^{q \hat e_0} \ .
\ee
Indeed,
\begin{eqnarray*}\label{eq:U_1-q_prod_}
\hat{U}_{1-q}^T(T)&=& e^{\hat e_0}\hat{u}_{0}e^{q(\hat e_{1}-\hat e_{0})}\hat{u}_{1}
\cdot\ldots\cdot \hat{u}_{K-1} e^{q(\hat e_0-\hat e_{K-1})}e^{-\hat e_0} \\
&=&
e^{\hat e_0} \hat{u}_{0}e^{-q\hat e_{0}} e^{q\hat e_1}\hat{u}_{1}e^{-q\hat e_{1}}
\cdot\ldots\cdot e^{q\hat e_{K-1}}\hat{u}_{K-1}e^{-q\hat e_{K-1}}e^{(q-1)\hat e_0}
 \\
&=&
e^{\hat e_0} \hat{u}_{0}e^{-q\hat e_{0}} \hat B_q e^{q\hat e_{K-1}}\hat{u}_{K-1}e^{-q\hat e_{K-1}}
\cdot\ldots\cdot e^{q\hat e_{1}}\hat{u}_{1}e^{-q\hat e_{1}}\hat B_q^{-1}e^{(q-1)\hat e_0} \\
&=&
e^{\hat e_0} \hat{u}_{0}(e^{-q\hat e_{0}} \hat B_q e^{q\hat e_{0}})\hat U_q(T)\hat u_0^{-1} (e^{-q\hat e_{0}} \hat B_q e^{q\hat e_{0}})^{-1}e^{-\hat e_0} = \hat C \hat U_q(T) \hat C^{-1}
\ .
\end{eqnarray*}

In this case, the symmetry $G_q = G_{1-q}$ is fulfilled automatically as the commutation~(\ref{eq:B_q_u0_commute})
leads to the common eigenbasis of both matrices $e^{-q \hat e_0}\hat B_q e^{q \hat e_0}$ and $\hat{u}_0$.
Thus,
the vectors
$\left|\vec{p}_{eq}(0)\ra$ and $\la\vec{1}\right|$
are the right and left eigenvectors of $e^{-q \hat e_0}\hat B_q e^{q \hat e_0}$, respectively, with the same eigenvalue $b$
\be\label{eq:B_q-symm_init_cond}
\left(e^{-q \hat e_0}\hat B_q e^{q \hat e_0}\right)\left|\vec{p}_{eq}(0)\ra = b \left|\vec{p}_{eq}(0)\ra, \quad \la\vec{1}\right|e^{-q \hat e_0}\hat B_q e^{q \hat e_0} = \la\vec{1}\right|b \
\ee
and this matrix can be diminished in (\ref{eq:G_q_solution}, \ref{eq:G_1-q_solution}) after the transformation (\ref{eq:B_q-symm}).
Note that the finite-time symmetry works even for lifted commutation relation~(\ref{eq:B_q_u0_commute}) if Eq.~(\ref{eq:B_q-symm_init_cond}) still holds. This hints that, in this concrete example, the periodic NESS fluctuation theorem is more restrictive on the drive than its finite-time counterpart.

As Eq.~(\ref{eq:B_q-symm}) works for all $k$ and for general step evolution operators,
the matrix $\hat B_q$ satisfies the following condition $\hat B_q^2=\hat I$ and thus all eigenvalues, including $b$ are $1$ or $-1$.
A reasonable example of the transformation $\hat B_q$ is the permutation of levels $E_n(t_k) = E_{P(n)}(t_{K-k})$ with $P(P(n))\equiv n$, leading, e.g., to anti-symmetric drive when in the second half period all the levels $E_n$ are put in the reversed order, $E_n(T-t)-E_m(T-t)=E_m(t)-E_n(t)$.
As discussed in the main text, this level permutation does not change the system itself, but affects the dynamics of the occupancies $p_n(t)$ and thus leads to some non-trivial dissipated work.

To sum up this section, we provide a particular example of the symmetry~(\ref{eq:U_1-q_sym}) being probably just the permutation of energy levels, which demonstrate that the finite-time fluctuation theorem can be less restrictive than its asymptotic counterpart.

\section{Three-step drive in two-level system}\label{App_Sec:3-step}
As mentioned in the main text for a two-level system, the only control parameter is $2\lambda(t)=\beta(E_1 - E_0)$.
Omitting the unimportant global energy shift one can take $E_0 = -E_1$ and
write the energy matrix in the form of Pauli matrix $\beta\hat E(t) = \sigma_z \lambda(t)$.
Then the free energy is $\beta F(t) = - \ln [2\cosh(\lambda(t)/2)]$, the equilibrium probability distribution vector
$\left|\vec{p}_{eq}(t)\ra = e^{-\sigma_z\lambda(t)}\left|\vec{1}\ra$
and the matrix of the tunneling rates reads as
%\be
%\Gamma_{01}(t) = \Gamma_0(t) e^{\lambda(t)}, \quad
%\Gamma_{10}(t) = \Gamma_0(t) e^{-\lambda(t)} \ .
%\ee
\be\label{eq:Gamma_n_mat}
\hat \Gamma(t)= \gamma(t)
\left(
  \begin{array}{rr}
-e^{\lambda(t)} & e^{-\lambda(t)}\\
e^{\lambda(t)} & -e^{-\lambda(t)}\\
  \end{array}
\right)
=\gamma(t)(\sigma_x-\hat I)e^{\sigma_z\lambda} \ .
\ee

The step evolution operator $\hat u_k = \exp[\hat{\Gamma}(t_{k}+0) \Delta t_k]$ can be written in a standard form (see, e.g., \cite{Esposito2013,BaratoChetrite2018})
\bea\label{eq:step_u(t)_TLS}
\hat u_k &=&
%e^{-\sigma_z\lambda_k/2}\left[\hat I + \frac{1-e^{-\Gamma_{\Sigma,k} \Delta t_k}}{2\cosh\lambda_k}\left(\sigma_x-e^{\sigma_z\lambda_k}\right)\right]e^{\sigma_z\lambda_k/2} \ ,
\hat I + \frac{1-e^{-|\gamma_{\min}(t_k)| \Delta t_k}}{2\cosh\lambda_k}\left(\sigma_x-\hat I\right)e^{\sigma_z\lambda_k}\nonumber\\
&=&\left|\vec{p}_{eq}(t_k)\ra\la\vec{1}\right|+e^{-|\gamma_{\min}(t_k)| \Delta t_k}\frac{\left(\hat I-\sigma_x\right)e^{\sigma_z\lambda_k}}{2\cosh\lambda_k}
 \ ,
\eea
with $-\gamma_{\min}(t_k) = 2\gamma_k\cosh\lambda_k>0$ and $\left|\vec{p}_{eq}(t_k)\ra\la\vec{1}\right| \equiv e^{-\sigma_z\lambda}\left(\hat I + \sigma_x\right)$.
The last line in (\ref{eq:step_u(t)_TLS}) confirms the general form~(\ref{eq:step_u(t)}), while
the first line for $\lambda = 0$ goes to (\ref{eq:step_u0(t)}).

According to Eq.~(\ref{eq:Delta_q_ln_ep_q}), the cumulative distribution function $\Delta_q(T)$ coincides with
the logarithm of the maximal eigenvalue of $\hat U_q(T)$. For the two-level system, this eigenvalue can be explicitly written (see, e.g., \cite{Esposito2013,BaratoChetrite2018})
\be\label{eq:ep_q}
\ep_q = \frac{  \tr \hat U_q(T) + \sqrt{ \left [ \tr \hat U_q(T)\right ]^{2} - 4 \det \hat U_q(T) }}{ 2} .
\ee
As follows from Eq.~(\ref{eq:det_U_q}), the determinant $\det \hat U_q(T)$ does not depend on $q$.
Thus using Eqs.~(\ref{eq:Delta_q_ln_ep_q}), (\ref{eq:G_q_tr_U_q}), and (\ref{eq:ep_q})
one concludes that the analysis of $\tr \hat U_q(T)$ is enough for both \rev{DFRs} (\ref{eq:Crooks+GC_in_G_q}).

Evaluating the trace of Eq.~(\ref{eq:U_q_prod}) one should keep only even powers of $\sigma_x$
\bea
\tr\hat U_q(T) =
%\tr\left[\prod_{k=0}^2 \left(\hat I - \frac{(1-e^{-\Gamma_{\Sigma,k}\Delta t_k})e^{\sigma_z \lambda_k}}{2\cosh\lambda_k}\right) \right]\nonumber\\
%+\sum_{k=0}^2\tr\left[ \sigma_x e^{\sigma_z(\lambda_k - \lambda_{k-1})x/2}\sigma_x \left(\hat I - \frac{(1-e^{-\Gamma_{\Sigma,k+1}\Delta t_{k+1}})e^{\lambda_{k+1} \sigma_z}}{2\cosh\lambda_{k+1}}\right)e^{-\sigma_z(\lambda_k - \lambda_{k-1})x/2}\right]\\
%\times\frac{(1-e^{-\Gamma_{\Sigma,k}\Delta t_k})}{2\cosh\lambda_{k}}\frac{(1-e^{-\Gamma_{\Sigma,k-1}\Delta t_{k-1}})}{2\cosh\lambda_{k-1}} = \\ =
%%
%B+
%\sum_{k=0}^2\tr\left[ e^{-\sigma_z 2(\lambda_k - \lambda_{k-1})x/2}\left(\hat I - \frac{(1-e^{-\Gamma_{\Sigma,k+1}\Delta t_{k+1}})e^{\lambda_{k+1} \sigma_z}}{2\cosh\lambda_{k+1}}\right)\right] = \\=
%B+
%\sum_{k=0}^2\left\{C_k\cosh\left[(\lambda_k - \lambda_{k-1})x\right]-S_k\sinh\left[(\lambda_k - \lambda_{k-1})x\right]\right\}
B+\sum_{k=0}^2&\bigl\{&C_k\cosh\left[(\lambda_k - \lambda_{k-1})(2q-1)\right]\nonumber\\&-&S_k\sinh\left[(\lambda_k - \lambda_{k-1})(2q-1)\right]\bigr\}
,
\eea
with
\bea
B &=& \tr\left[\prod_{k=0}^2 \left(\hat I - \frac{(1-e^{-|\gamma_{\min}(t_k)|\Delta t_k})e^{\sigma_z \lambda_k}}{2\cosh\lambda_k}\right) \right]\ ,\\
C_k &=& (1+e^{-|\gamma_{\min}(t_{k+1})|\Delta t_{k+1}})\frac{(1-e^{-|\gamma_{\min}(t_k)|\Delta t_k})}{2\cosh\lambda_{k}}\frac{(1-e^{-|\gamma_{\min}(t_{k-1})|\Delta t_{k-1}})}{2\cosh\lambda_{k-1}} \ ,\\
S_k &=& 2\sinh\lambda_{k+1}\frac{(1-e^{-|\gamma_{\min}(0)|\Delta t_0})}{2\cosh\lambda_{0}}\frac{(1-e^{-|\gamma_{\min}(t_1)|\Delta t_1})}{2\cosh\lambda_{1}}\frac{(1-e^{-|\gamma_{\min}(t_2)|\Delta t_2})}{2\cosh\lambda_{2}} \ .
\eea
Here, the indices $k$ are considered modulo $K=3$.

Thus, the symmetry $\tr \hat U_q(T) = \tr \hat U_{1-q}(T)$ is valid, if and only if, for any $q$
\be\label{eq:3step_cond_sinh}
\sum_{k=0}^2 \sinh\left[(\lambda_k - \lambda_{k-1})(2q-1)\right]\sinh\lambda_{k+1} = 0 \ .
\ee

Without loss of generality, let's consider $\lambda_0<\lambda_1<\lambda_2$ and send $q\to\infty$.
Then one concludes that the coefficient $\sinh \lambda_1$ in front of the hyperbolic sine with the largest increment
$\lambda_2-\lambda_0>\lambda_1-\lambda_0,\lambda_2-\lambda_1>0$ should go to zero, thus, $\lambda_1=0$.
As a result, Eq.~(\ref{eq:3step_cond_sinh}) reduces to
\be
\sinh\left[\lambda_0\right]\sinh\left[\lambda_2 (2q-1)\right]-\sinh\left[\lambda_2\right]\sinh\left[\lambda_0 (2q-1)\right] = 0,
\ee
and leads to $\lambda_0=-\lambda_2$.
This completes the proof of Eq.~(\ref{eq:3step_cond}).

\section{Relations~(\ref{eq:G_q_tr_U_q}, \ref{eq:G_q_2_tr_U_q}) between $G_q$ and $\tr U_q$}\label{App_Sec:G_q_trU_q}
As the step evolution operator $\hat u_k(\Delta t_k)$ entering the expression~(\ref{eq:U_q_prod}) for the total evolution operator
has the only non-negative eigenvalue $0$ corresponding to the left $\la \vec{1}\right|$ and right $\left| \vec{p}_{eq}\ra$ eigenvectors,
it can be represented in the form~(\ref{eq:step_u(t)})
\be\label{eq:step_u(t)_general}
\hat u_k(\Delta t_k) = \left|\vec{p}_{eq}(t_k)\ra\la \vec{1}\right| + \hat{\delta u}_k (\Delta t_k) \ ,
\ee
with the elements of the matrix $\hat{\delta u}_k (\Delta t_k)$ exponentially decaying with the time duration $\Delta t_k$.
As a result,
\be\label{eq:U_q_prod_}
\lim\limits_{\Delta t_0\to\infty}\hat{U}_q(T)= e^{q(\hat e_{K-1}-\hat e_0)}\hat{u}_{K-1} e^{q(\hat e_{K-2}-\hat e_{K-1})}\hat{u}_{K-2}\cdot\ldots\cdot e^{q(\hat e_0-\hat e_1)}\left|\vec{p}_{eq}(0)\ra\la \vec{1}\right| \
\ee
and thus the trace of the latter $\lim_{\Delta t_0\to \infty} \tr \hat{U}_q(T)$ coincides with the expression for $G_{q}(T)$ (\ref{eq:G_q_tr_U_q}) and this concludes the derivation.

For the two-level system the expression (\ref{eq:step_u(t)_general}) simplifies to (\ref{eq:step_u(t)_TLS})
and thus
\be\label{eq:U_q=G_q+du_0}
\tr U_q(T,\Delta t_0) = G_q +
e^{-|\gamma_{\min}(0)|\Delta t_0}\tr\left[e^{q(\hat e_{K-1}-\hat e_0)}\hat{u}_{K-1} \cdot\ldots\cdot \hat{u}_1 e^{q(\hat e_0-\hat e_1)} \hat{\delta u}_0\right] \ .
\ee

Using the latter expression~(\ref{eq:U_q=G_q+du_0}) for $\tr U_q(T,\Delta t_0)$ and $\tr U_q(T,\Delta t_0')$
and excluding the second terms from them, one comes to Eq.~(\ref{eq:G_q_2_tr_U_q}).
The more general expression for $\tr U_q(T,\Delta t_0'')$ via $\tr U_q(T,\Delta t_0)$ and $\tr U_q(T,\Delta t_0')$ takes the form
\bea
%\tr U_q(T,\Delta t_0'') = \frac{\left(1-e^{-\Gamma_{\Sigma,0}(\Delta t_0''-\Delta t_0')}\right)\tr \hat U_q(T, \Delta t_0)-\left(e^{-\Gamma_{\Sigma,0}(\Delta t_0-\Delta t_0')}-e^{-\Gamma_{\Sigma,0}(\Delta t_0''-\Delta t_0')}\right)\tr \hat U_q(T, \Delta t_0')}
%{1-e^{-\Gamma_{\Sigma,0}(\Delta t_0-\Delta t_0')}} \ .
\tr U_q(T,\Delta t_0'') &=& \frac{\left(e^{-|\gamma_{\min}(0)|\Delta t_0'}-e^{-|\gamma_{\min}(0)|\Delta t_0''}\right)\tr \hat U_q(T, \Delta t_0)}
{e^{-|\gamma_{\min}(0)|\Delta t_0'}-e^{-|\gamma_{\min}(0)|\Delta t_0}}\nonumber \\&-&
                                \frac{\left(e^{-|\gamma_{\min}(0)|\Delta t_0}-e^{-|\gamma_{\min}(0)|\Delta t_0''}\right)\tr \hat U_q(T, \Delta t_0')}
{e^{-|\gamma_{\min}(0)|\Delta t_0'}-e^{-|\gamma_{\min}(0)|\Delta t_0}} \ .
\eea


\smallskip
\section*{References}
\begin{thebibliography}{99}

\bibitem{BK}
G. N. Bochkov and Yu. E. Kuzovlev, JETP {\bf45}, 125 (1977).

\bibitem{BK-review}
G. N. Bochkov and Yu. E. Kuzovlev,
Physics-Uspekhi {\bf56}, 590 - 602 (2013).

\bibitem{Evans-Cohen1993}
D. J. Evans, E. G. D. Cohen, and G. P. Morriss,
Phys. Rev. Lett. {\bf71}, 2401 (1993).

\bibitem{GC1995}
G.~Gallavotti and E.~G.~D.~Cohen,
Phys. Rev. Lett. {\bf74}, 2694 (1995).

\bibitem{Kurchan1998}
J. Kurchan, %Fluctuation theorem for stochastic dynamics
J. Phys. A: Math. Gen. {\bf31}, 3719 (1998).

\bibitem{Lebowitz-Spohn1999}
J. L. Lebowitz and H. Spohn,
J. Stat. Phys. {\bf95}, 333 (1999).

\bibitem{Evans-Searles1994}
D. J. Evans and D. J. Searles, %Equilibrium microstates which generate second law violating steady states
Phys. Rev. E {\bf50}, 1645 (1994).

\bibitem{Jarzynski1997}
C.\ Jarzynski, Phys. Rev. Lett. {\bf 78}, 2690 (1997).

\bibitem{Crooks1999}
G. Crooks, Phys. Rev. E {\bf 60}, 2721 (1999).

\bibitem{Tietz-Seifert2006}
C. Tietz, S. Schuler, T. Speck, U. Seifert, and J. Wrachtrup,
Phys. Rev. Lett. {\bf97}, 050602 (2006).

\bibitem{Shargel-Chou2009}
B. H. Shargel and T. Chou, %Fluctuation theorems for entropy production and heat dissipation in periodically driven Markov chains
J. Stat. Phys. {\bf137}, 16588 (2009).

\bibitem{Seifert2005}
U. Seifert, Phys. Rev. Lett. {\bf 95}, 040602 (2005).

\bibitem{Garcia-Garcia2010}
R. Garcia-Garcia, D. Dominguez, V. Lecomte, and A. B. Kolton,
%Unifying approach for fluctuation theorems from joint probability distributions
Phys. Rev. E {\bf82} 030104(R) (2010).

\bibitem{Seifert2012}
U. Seifert, Rep. Prog. Phys. {\bf75}, 126001 (2012).

\bibitem{Liphardt2002}
J. Liphardt, S. Dumont, S. B. Smith, I. Tinoco Jr., and
C. Bustamante, %Equilibrium information from nonequilibrium measurements in an experimental test of Jarzynski's equality
Science {\bf296}, 1832 (2002).

\bibitem{Evans2002}
G. M. Wang, E. M. Sevick, E. Mittag, D. J. Searles, and D. J. Evans,
Phys. Rev. Lett. {\bf89}, 050601 (2002).

\bibitem{Ciliberto2005}
F. Douarche, S. Ciliberto, A. Petrosyan and I. Rabbiosi,
Europhys. Lett. {\bf70}, 593 (2005).

\bibitem{Hoang2018}
T. M. Hoang, R. Pan, J. Ahn, J. Bang, H. T. Quan, and T. Li,
Phys. Rev. Lett. {\bf120}, 080602 (2018).

\bibitem{Bustamante2004}
E. H. Trepagnier, C. Jarzynski, F. Ritort, G. E. Crooks,
C. J. Bustamante, and J. Liphardt, %Experimental test of Hatano and Sasa's nonequilibrium steady-state equality
Proc. Natl Acad. Sci. USA {\bf101}, 15038 (2004).

\bibitem{Collin2005}
D. Collin, F. Ritort, C. Jarzynski, S. B. Smith, I. Tinoco, and
C. Bustamante, %Verification of the Crooks fluctuation theorem and recovery of RNA folding free energies
Nature {\bf437}, 231 (2005).

\bibitem{Ciliberto2005_Resistor}
N. Garnier and S. Ciliberto,
Phys. Rev. E {\bf71}, 060101(R) (2005).

\bibitem{Schuler-Seifert2005}
S. Schuler, T. Speck, C. Tietz, J. Wrachtrup, and U. Seifert,
Phys. Rev. Lett. {\bf94}, 180602 (2005).

\bibitem{Saira2012}
O.-P. Saira, Y. Yoon, T. Tanttu, M. M{\"o}tt{\"o}nen, D.V. Averin, and J. P. Pekola,
Phys. Rev. Lett. {\bf109}, 180601 (2012).

\bibitem{Roldan-Ciliberto2015}
L. Granger, J. Mehlis, {\'E} Rold{\'a}n, S. Ciliberto, and H. Kantz,
New J. Phys. {\bf17}, 065005 (2015).

\bibitem{Hofmann2015}
A. Hofmann, V. F. Maisi, C. R{\"o}ssler, J. Basset, T. Kr{\"a}henmann, P. M{\"a}rki, T. Ihn, K. Ensslin, C. Reichl, W. Wegscheider,
Phys. Rev. B 93, 035425 (2016).

\bibitem{Koski2013}
J. V. Koski, T. Sagawa, O-P. Saira, Y. Yoon, A. Kutvonen, P. Solinas, M. M{\"o}tt{\"o}nen,
T. Ala-Nissil{\"a}, and J. P. Pekola,
Nat. Phys. {\bf9}, 644 (2013).

\bibitem{Serra2014}
T. Batalh$\tilde{\rm a}$o, A. M. Souza, L. Mazzola, R. Auccaise, I. S. Oliveira, J. Goold, G. De Chiara, M. Paternostro, and R. M. Serra,
%Experimental reconstruction of work distribution and verification of fluctuation relations at the full quantum level
Phys. Rev. Lett. {\bf113}, 140601 (2014).

\bibitem{Nakamura2010}
S. Nakamura, Y. Yamauchi, M. Hashisaka, K. Chida, K. Kobayashi, T. Ono, R. Leturcq, K. Ensslin, K. Saito, Y. Utsumi, and A. C. Gossard
Phys. Rev. Lett. {\bf104}, 080602 (2010).

\bibitem{Gasparinetti2015}
S. Gasparinetti, K. L. Viisanen, O.-P. Saira, T. Faivre, M. Arzeo, M. Meschke, and J. P. Pekola,
%Fast Electron Thermometry for Ultrasensitive Calorimetric Detection,
Phys. Rev. Appl. {\bf 3}, 014007 (2015).

\bibitem{Feshchenko2015}
A. V. Feshchenko, L. Casparis, I. M. Khaymovich, D. Maradan, O.-P. Saira, M. Palma, M. Meschke, J. P. Pekola, D. M. Zumb\"uhl
%Tunnel junction thermometry down to millikelvin temperatures
Phys. Rev. Appl. {\bf4}, 034001 (2015).

\bibitem{Koski2015_AutoMD}
J. V. Koski, A. Kutvonen, I. M. Khaymovich, T. Ala-Nissil{\"a}, and J. P. Pekola,
Phys. Rev. Lett. {\bf115}, 260602 (2015).

\bibitem{Saira2016}
O.-P. Saira, M. Zgirski, K.L. Viisanen, D.S. Golubev, and J.P. Pekola
%Dispersive thermometry with a Josephson junction coupled to a resonator
Phys. Rev. Appl. {\bf6}, 024005 (2016).

\bibitem{Zgirski2018}
M. Zgirski, M. Foltyn, A. Savin, M. Meschke, and J. Pekola,
%Nanosecond thermometry with Josephson junction
Phys. Rev. Applied {\bf10}, 044068 (2018).

\bibitem{Wang2018}
L. Wang, O.-P. Saira, J. P. Pekola,
%Fast thermometry with a proximity Josephson junction
Appl. Phys. Lett. {\bf112}, 013105 (2018)

\bibitem{Sagawa-Ueda2010}
T. Sagawa and M. Ueda, %Generalized Jarzynski equality under nonequilibrium feedback control
Phys. Rev. Lett. {\bf104}, 090602 (2010).

\bibitem{Sagawa-Ueda2012}
T. Sagawa and M. Ueda, %Nonequilibrium thermodynamics of feedback control
Phys. Rev. E {\bf85}, 021104 (2012).

\bibitem{Potts2018}
P. P. Potts and P. Samuelsson,
Phys. Rev. Lett. {\bf121}, 210603 (2018).

\bibitem{Landauer1961}
R. Landauer, IBM J. Res. Develop. {\bf5}, 183 (1961).

\bibitem{Landauer1988}
R. Landauer, Nature {\bf335}, 779 (1988).


\bibitem{Toyabe2010}
S. Toyabe, T. Sagawa, M. Ueda, E. Muneyuki, and M. Sano,
Nature Phys. {\bf6}, 988 (2010).

\bibitem{Roldan2014}
E. Roldan, I. A. Martinez, J. M. R. Parrondo, and D. Petrov,
Nature Phys. {\bf 10}, 457 (2014).

\bibitem{Koski2014_MD_PNAS}
J. V. Koski, V. F. Maisi, J. P. Pekola, and D. V. Averin,
Proc. Natl. Acad. Sci. {\bf111}, 13786 (2014).

\bibitem{Koski2014_MD_PRL}
J. V. Koski, V. F. Maisi, T. Sagawa, and J. P. Pekola,
Phys. Rev. Lett. {\bf113}, 030601 (2014).

\bibitem{Vidrighin2016}
M. D. Vidrighin, O. Dahlsten, M. Barbieri, M. S. Kim, V. Vedral, and I. A. Walmsley,
Phys. Rev. Lett. {\bf116}, 050401 (2016).

\bibitem{Ribezzi}
M. Ribezzi-Crivellari, F. Ritort et al.
private communications.

\bibitem{Chida2015}
K. Chida, K. Nishiguchi, G. Yamahata, H. Tanaka, and A. Fujiwara,
Appl. Phys. Lett. {\bf107}, 073110 (2015).

\bibitem{Wagner2016}
T. Wagner, P. Strasberg, J. C. Bayer, E. P. Rugeramigabo, T. Brandes, and R.J. Haug,
Nature Nanotech. {\bf 12}, 218-222 (2017).

\bibitem{Pekola2015_NatPhys_Review}
J. P. Pekola,
Nat. Phys. {\bf11}, 118 (2015).

\bibitem{Pekola2019_AnnRev}
J. P. Pekola and I. M. Khaymovich,
%Thermodynamics in Single-Electron Circuits and Superconducting Qubits,
Annu. Rev. Condens. Matter Phys. {\bf10}, 193 (2019).

\bibitem{Chetrite_Gupta}
R. Chetrite, S. Gupta,
J. Stat. Phys. {\bf143}, 543 (2011).

\bibitem{Neri_PRX}
I. Neri, {\'E}. Rold{\'a}n, F. J{\"u}licher,
Phys. Rev. X {\bf7}, 011019 (2017).

\bibitem{Pigolotti_PRL}
S. Pigolotti, I. Neri, {\'E}. Rold{\'a}n, F. J{\"u}licher,
Phys. Rev. Lett. {\bf119}, 140604 (2017).

\bibitem{Neri2019}
I. Neri, {\'E}. Rold{\'a}n, S. Pigolotti, F. J{\"u}licher,
%Integral Fluctuation Relations for Entropy Production at Stopping Times,
arxiv:1903.08115 (2019).

\bibitem{Singh_Sinf}
I.	S. Singh, {\'E}. Rold{\'a}n, I. Neri, I. M. Khaymovich, D. S. Golubev, V. F. Maisi, J. T. Peltonen, F. J{\"u}licher, and J. P. Pekola,
%"Records of entropy production in an electronic double dot",
arXiv:1712.01693 (2017).

\bibitem{Manzano2019}
G. Manzano, R. Fazio, {\'E}. Rold{\'a}n,
%Quantum Martingale Theory and Entropy Production
arxiv:1903.02925 (2019).

\bibitem{Chetrite_Q-Martingale}
R. Chetrite, S. Gupta, I. Neri, and {\'E}. Rold{\'a}n,
%Martingale theory for housekeeping heat
arXiv:1810.09584 (2018).

\bibitem{VerleyLacoste2012}
G. Verley and D. Lacoste,
Phys. Rev. E {\bf86}, 051127 (2012).

\bibitem{Sekimoto}
K.~Sekimoto,  {\it Stochastic energetics}, Lecture Notes in Physics {\bf799} (Springer, 2010).

\bibitem{Singh_Thesis}
Shilpi Singh, PhD thesis, Aalto University School of Science (2019).
\url{http://urn.fi/URN:ISBN:978-952-60-8420-6}


\bibitem{Singh_1st_passage}
S.~Singh, P.~Menczel, D. S. Golubev, I. M. Khaymovich, J. T. Peltonen, C. Flindt, K. Saito, {\'E}. Rold{\'a}n, J. P. Pekola,
arXiv:1809.06870 (2018).

\bibitem{MF_analogy}
I.~M.~Khaymovich, J.~V.~Koski, O.-P.~Saira, V.~E.~Kravtsov, J.~P.~Pekola, Nat. Comm. {\bf6}, 7010 (2015).

\bibitem{Touchette2009}
H. Touchette, Phys. Rep. {\bf478}, 1 (2009).

\bibitem{CuetaraEsposito2014}
G. B. Cuetara, M. Esposito, and A. Imparato,
Phys. Rev. E {\bf89}, 052119 (2014).

\bibitem{RaoEsposito2018}
R. Rao and M. Esposito,
arXiv:1807.09242 (2018).

\bibitem{Talkner1999}
P. Talkner, New J. Phys. {\bf1}, 4 (1999).

\bibitem{Bagrets_Nazarov2003}
D. A. Bagrets, Yu. V. Nazarov, Phys. Rev. B {\bf 67}, 085316 (2003).

\bibitem{Singh2016}
S. Singh, J. T. Peltonen, I. M. Khaymovich, J. V. Koski, C. Flindt, and J. P. Pekola,
Phys. Rev. B {\bf94}, 241407(R) (2016).

\bibitem{Esposito2013}
G. Verley, C. Van den Broeck, and M. Esposito Phys. Rev. E {\bf 88}, 032137 (2013).

\bibitem{BaratoChetrite2018}
A. C. Barato and R. Chetrite,
J. Stat. Mech {\bf2018}, 053207 (2018).


\bibitem{Breuer_Petruccione_book}
H.-P. Breuer and F. Petruccione, {\it The Theory of Open Quantum Systems}, (Oxford University Press, 2002).

\end{thebibliography}
\end{document}